\documentclass[12pt]{iopart}

\usepackage{iopams}
\usepackage{epsfig,subfigure}
\usepackage{rotating}
\usepackage{xcolor}
\newcommand{\msun}{M_\odot}
\newcommand{\be}{\begin{equation}}
\newcommand{\ee}{\end{equation}}
\newcommand{\bea}{\begin{eqnarray}}
\newcommand{\eea}{\end{eqnarray}}

\begin{document}

\title[]{Improvement of the parameter measurement accuracy by the third-generation gravitational wave detector Einstein Telescope}

\author{Hee-Suk Cho}
\ead{chohs1439@pusan.ac.kr}
\address{Department of Physics, Pusan National University\\
Busan, 46241, Korea\\
Extreme Physics Institute, Pusan National University\\
Busan, 46241, Korea}

\begin{abstract}
The Einstein Telescope (ET) has been proposed as one of the third-generation gravitational wave (GW) detectors.
The sensitivity of ET would be a factor of 10 better than the second-generation GW detector, Advanced LIGO (aLIGO); thus, the GW source parameters could be measured with much better accuracy. In this work, we show how the precision in parameter estimation can be improved between aLIGO and ET by comparing the measurement errors.
We apply the TaylorF2 waveform model defined in the frequency domain to the Fisher matrix method which is a semi-analytic approach for estimating GW parameter measurement errors.
We adopt as our sources low-mass binary black holes with the total masses of $M\leq 16\msun $ and the effective spins of $-0.9 \leq \chi_{\rm eff} \leq 0.9$ and calculate the measurement errors of the mass and the spin parameters using $10^4$ Monte-Carlo samples randomly distributed in our mass and spin parameter space.
We find that for the same sources ET can achieve $\sim 14$ times better signal-to-noise ratio than aLIGO
and the error ratios ($\sigma_{\lambda, \rm ET}/\sigma_{\lambda, \rm aLIGO}$) for the chirp-mass, symmetric mass ratio, and effective spin parameters can be lower than $7\%$ for all binaries.
We also consider the equal-mass binary neutron stars with the component masses of 1, 1.4, and 2 $\msun$ and find that the error ratios for the mass and the spin parameters can be lower than $1.5 \%$.
In particular, the measurement error of the tidal deformability $\tilde{\Lambda}$ can also be significantly reduced by ET, with the error ratio of $3.6 - 6.1 \%$.
We investigate the effect of prior information by applying the Gaussian prior on the coalescence phase $\phi_c$ to the Fisher matrix and find that 
the error of the intrinsic parameters can be reduced to $\sim 70\%$ of the original priorless error ($\sigma_{\lambda}^{\rm priorless}$) if the standard deviation of the prior is similar to $\sigma_{\phi_c}^{\rm priorless}$.

\end{abstract}


\section{Introduction}
In the past few years, a new window to see the universe has been opened
through the successful observations of the gravitational waves (GWs) 
by the ground-based GW detectors, Advanced LIGO (aLIGO) \cite{ALIGO} and Advanced Virgo \cite{AVirgo}.
So far, LIGO and Virgo have conducted a few times science observations successfully
and they are currently preparing the next operation with upgraded detector performance.
The source of the first detected GW signal, named GW150914,  was a merging binary black hole (BBH)
with the masses of $\sim 36 \msun$ and $\sim 29 \msun$ \cite{GW150914}.
GW150914 was not only the first direct
detection of GWs but also the first observation of a BBH merger.
This demonstrated the existence of stellar-mass
black holes more massive than $\sim25\msun$.
On the other hand, a GW signal  
whose origin was a merging binary neutron star (BNS) was detected on 17 August 2017; thus, named GW170817. \cite{GW170817}.
At the same time, the detection of the gamma-ray burst GRB170817A and the electromagnetic follow-up
were made successfully with respect to the same source of GW170817 \cite{GW170817GRB,GW170817EM}.
Consequently, this event could begin a new era of multi-messenger astronomy.

The efficiency of GW searches mainly depends on the detector sensitivity, so 
 the GW observatories have continuously upgraded the instrument performance.
The first-generation GW detectors, initial LIGO and initial Virgo, were able to observe the GWs
only in the frequency range from $40$ Hz to a few hundred Hz, and did not detect any GW signals.
The second-generation GW detectors, aLIGO and Advanced Virgo, could lower the 
 low-frequency limit down to $10$--$20$ Hz and consequently have detected  GW signals successfully.  
 About 50 GW signals have been observed so far, 
all of them were originated from merging compact binaries consisting of a black hole and/or a neutron star \cite{GWTC-1,GWTC-2}.

After detection of a GW signal, the parameters of the source, such as the mass, the spin, the distance, the sky position, etc.,
can be measured by the parameter estimation analysis.
The measurement accuracy of each parameter depends on 
how the parameter contributes to the description of the binary motion and consequently, to the shape of the waveforms.
The leading and the next-to-leading contributions to the waveforms are the mass and the spin parameters, respectively;
thus, those can be mainly measurable in GW parameter estimation. 
For the case GW150914, the $90 \%$ symmetric credible intervals are 10--20 \% for the component masses 
and  0.14--0.15 for the effective spin $\chi_{\rm eff}$ \cite{GW150914PE1,GW150914PE2}.
On the other hand, the neutron star tidal effect can be incorporated in the post-Newtonian (PN) waveforms.
However, since the tidal correction terms appear from the 5PN order,
its effect is relatively very small compared to that of the mass or the spin. 
For the case GW170817, assuming the low-spin prior ($|\chi_i| \leq 0.05$), the one-sided $90 \%$ lower or upper limits for the component masses are $(1.36, 1.60) \msun$ and $(1.16, 1.36) \msun$, the effective spin is  $0.00^{+0.02}_{-0.01}$ (where $x^{+z}_{-y}$ represent the median, $5\%$ lower limit, and $95\%$ upper limit),
and the $90\%$ symmetric credible intervals for the tidal deformability $\tilde {\Lambda}$ is $300^{+500}_{-190}$ \cite{GW170817PE}.
Although the tidal deformability was poorly measured, that result could roughly constrain various theoretical models on the neutron star equation of state \cite{GW170817PE,GW170817EOS}.

\begin{figure}[t]
\begin{center}
\includegraphics[width=10cm]{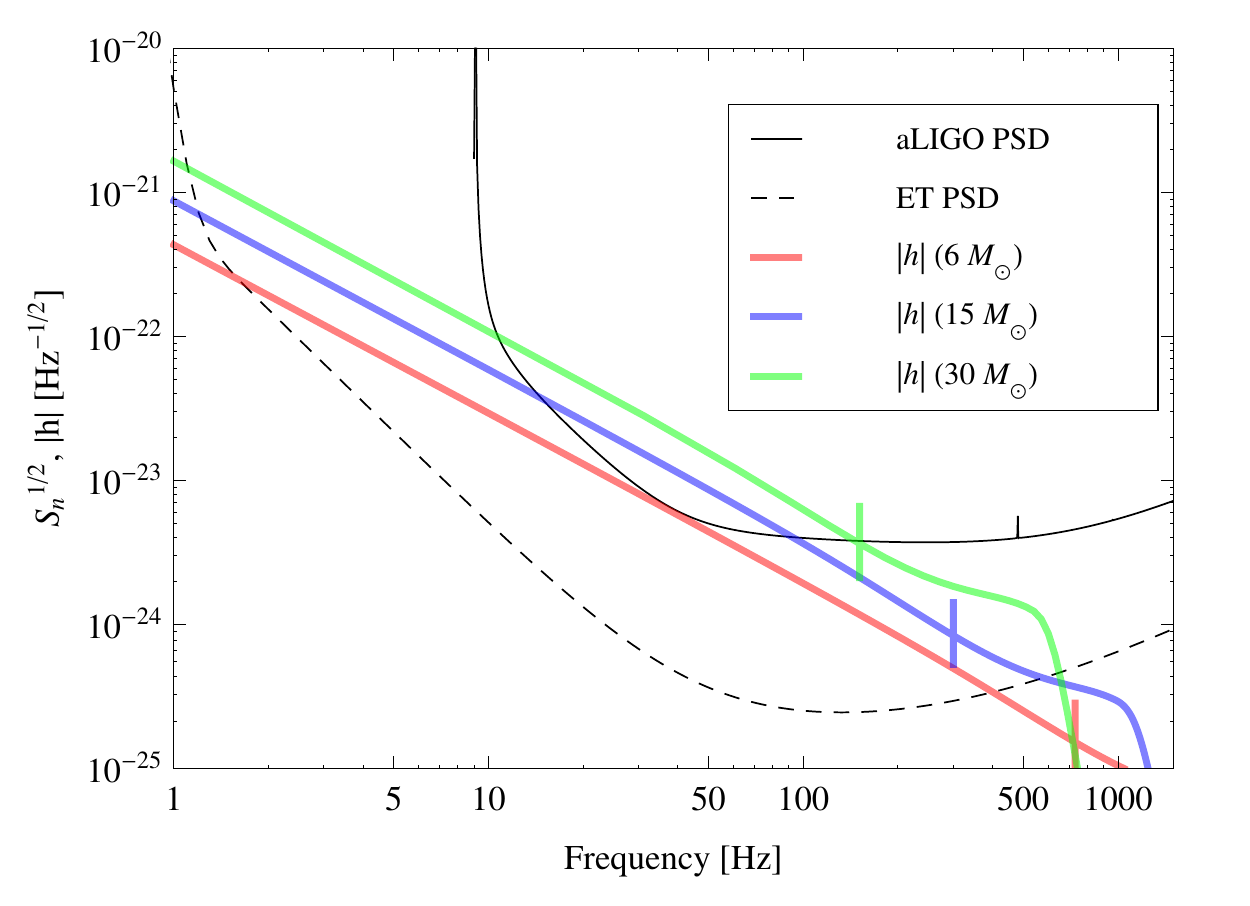}
\caption{Detector sensitivity curves for aLIGO and ET and amplitudes $|h|$ of the inspiral-merger-ringdown waveforms for BBHs with the total masses of $6, 15$ and $30 \msun$ calculated by using the inspiral-merger-ringdown waveform model (IMRPhenomD) \cite{Kha16}. The effective distance is 400 Mpc. The vertical line of each waveform indicates the frequency of the innermost stable circular orbit.  \label{fig.psd}}
\end{center}
\end{figure}

The Einstein Telescope (ET) is conceived to be one of the third-generation GW detectors.
ET will achieve a greatly improved sensitivity by increasing the size of the interferometer from the 3km arm length of the Virgo detector to 10km, and by implementing a series of new technologies \cite{ETwebsite}.
The sensitivity of ET would be a factor of 10 better than the second-generation detectors and it could
also extend the low-frequency limit down to 1--3 Hz.
Thus, it is expected that the measurement accuracy of GW source parameters could be much improved. 
In particular, due to the wide frequency band, ET could observe the BNS inspiral phase and the onset of tidal effects with a high signal-to-noise ratio (SNR).
Figure~\ref{fig.psd} shows the detector sensitivity curves for ET and aLIGO.
We consider the zero-detuned, high-power noise power-spectral-density (PSD) for aLIGO  \cite{aLIGOpsd}
and one of the design noise PSD curves for ET \cite{ETpsd}. 
Here, the ET curve is obtained by using the analytical fit to the ET-B noise PSD given by \cite{ETpsd-fit}
\be
S_n(f)=10^{-50} (2.39\times 10^{-27} x^{-15.64}+0.349 x^{-2.145}+1.76 x^{-0.12}+0.409 x^{1.10}),
\ee 
where $x=f/100$ Hz.

We use the TaylorF2 waveform model which is an analytic function expressed in the frequency domain.
This model is the most simple waveform model and has been widely used in GW data analysis studies.
Typically, the TaylorF2 waveform terminates at the frequency of the innermost stable circular orbit (ISCO)
and describes only the inspiral part of the inspiral-merger-ringdown waveforms for BBHs.
As the binary mass decreases, the ISCO frequency becomes higher,
consequently, the merger-ringdown contribution to the wave phasing becomes smaller. 
Therefore, TaylorF2 is suitable only for a sufficiently low-mass binary system in which 
the merger-ringdown contribution is negligible compared to the inspiral contribution.
In figure~\ref{fig.psd}, for example,  we depict the amplitude spectra of the waveforms as well as the ISCO frequencies for BBHs with masses of $6, 15,$ and $30 \msun$.
Several authors have investigated the effectualness of the inspiral-only template models with respect to the inspiral-merger-ringdown signals \cite{Far09,Bro13,Buo09,Aji08,Cho15a,Cho15b}.
They pointed out that the total mass should be lower than $\sim 15\msun$ to achieve the fitting factors over $\sim 0.97$, which corresponds to a loss of detection rates of $\sim 10\%$.

We calculate the measurement errors of the source parameters using both the aLIGO and the ET detector sensitivities
and show how the measurement accuracy can be improved with ET. 
We adopt the Fisher matrix method which is a semi-analytic approach for estimating the measurement errors of the GW source parameters. 
The Fisher matrix approximation may not represent the results of a full parameter estimation analysis due to several well-known limitations.
For example, since the Fisher matrix is constructed assuming a high SNR, it is only valid at high SNR limits.
In addition, there is no simple way to incorporate non-Gaussian priors into the Fisher matrix posterior function,
especially if the parameter being considered has physical boundaries, such as the spin parameter ($|\chi_i|\leq 1$). 
The authors of \cite{PhysRevD.88.084013} have also found that the disagreement between the Fisher Matrix method and the Markov-chain Monte Carlo (MCMC) simulations
increases with total mass.
Later, the authours of \cite{Mandel_2014} and \cite{Cho_2014} have found that such disagreement is more likely to occur with higher-mass binaries, 
where the inspiral-only waveform abruptly terminates within the detector frequency band.
This is also a major cause of the loss of fitting factor in search analysis as discussed above.
In this work, therefore, we only consider the BNSs and the low-mass BBHs with the total masses of $M\leq 16\msun $. 
We use the geometrized unit $G=c=1$.


\section{Waveform model : TaylorF2}\label{sec2}
In the GW data analysis, TaylorF2 is the most commonly used waveform model because this model is expressed simply by the post-Newtonian (pN) series expansions in closed form \cite{Sat91,Cut94,Poi95}. 
The TaylorF2 waveform can be obtained from the time-domain waveform via the stationary phase approximation
and the wave function can be expressed in the frequency domain as
\be \label{eq.TaylorF2}
h(f)={M_{\rm c}^{5/6} \over \pi ^{2/3} D_{\rm eff} } \sqrt{{5 \over 24}} f^{-7/6} e^{i \Psi(f)},
\ee
where $M_{\rm c}=\eta^{3/5}M$ (where $M=m_1+m_2$ is the total mass of the binary and $\eta=m_1m_2/M^2$ is the symmetric mass ratio) is the chirp mass
and $D_{\rm eff}$ is the effective distance of the source. 
The effective distance is determined by the extrinsic parameters, those are the true distance of the
source and several geometrical factors that relate the source
orientation to the detector orientation \cite{All12}. 
In a single-detector analysis, the only well-constrained extrinsic factor is $D_{\rm eff}$,
if we include $D_{\rm eff}$ in the Fisher matrix, we can also calculate the error of $D_{\rm eff}$.
However, the correlations between $D_{\rm eff}$ and the intrinsic parameters are negligible, 
so the inclusion of $D_{\rm eff}$ does not change the measurement errors of the intrinsic parameters (see, table \ref{tab.5dvs6d}).
Therefore, we do not take into account the extrinsic parameters in our analysis
by choosing a fixed effective distance.
We show the effect of the choice of various extrinsic parameters on the measurement errors of the intrinsic parameters in \ref{ap.extrinsic}.
In addition, we only consider an aligned-spin binary system, in which the binary does not precess.

The wave phase of TaylorF2 is given as
\be \label{eq.Psi}
\Psi(f)=2\pi ft_c -2 \phi_c -{\pi\over 4} + {3 \over 128 \eta v^5} [\phi_{\rm 3.5pN}(f)+ \phi^{\rm Tidal}(f)],
\ee
where $v= (\pi f M)^{1/3}$ is the post-Newtonian expansion parameter, 
$t_c$ and $\phi_c$ are the coalescence time and the phase at coalescence instant.
The 3.5PN order phase equation is expressed as~\cite{Aru09}
\bea   \label{eq.SPA_PP}
\phi_{\rm 3.5pN} = 1 &+&\left( \frac{3715}{756} + \frac{55}{9}\eta \right)v^2 +(4\beta- 16\pi) v^3 \nonumber \\
&+& \left( \frac{15293365}{508032} + \frac{27145}{504}\eta + \frac{3085}{72}\eta^2 -10 \sigma  \right) v^4  \nonumber\\ 
&+& \left(\frac{38645 \pi}{756} - \frac{65 \pi }{9}\eta  -\gamma \right)(1  + 3\log v) v^5   \nonumber   \\
&+& \bigg\{ \frac{11583231236531}{4694215680} - \frac{640}{3}\pi^2 -\frac{6848\gamma_{\rm E} }{21} -\frac{6848 \log\left(4{v}\right)}{ 21}   \nonumber\\ 
&&+  \left( \frac{2255{\pi }^2}{12}  - \frac{15737765635}{3048192}\right)\eta +\frac{76055}{ 1728}\eta^2-\frac{127825}{ 1296}\eta^3\bigg\} v^6  \nonumber  \\
&+& \left(\frac{77096675 \pi}{254016} + \frac{378515\pi}{1512}\eta 
- \frac{74045\pi}{756}\eta^2\right)v^7,
\eea
where $\gamma_{\rm E}=0.577216...$ is the Euler constant, and 
the terms $\beta, \sigma,$ and $\gamma$ denote the leading-order spin-orbit coupling, leading-order spin-spin coupling, and next-to-leading-order spin-orbit
coupling, respectively \cite{Aji11}:
\bea
\beta &=& \frac{113}{12}\left(\chi_s+\delta \chi_a -\frac{76 \eta}{113}\chi_s\right),  \nonumber\\
 \sigma&=&\chi_a^2\left(\frac{81}{16}-20 \eta\right)+\frac{81 \chi_a \chi_s \delta}{8} + \chi_s\left(\frac{81}{16}  -\frac{\eta}{4}\right),   \nonumber\\
\gamma&=&\chi_a \delta \left(\frac{140 \eta}{9}+\frac{732985}{2268}\right) + \chi_s \left(\frac{732 985}{2268}-\frac{24 260 \eta}{81}-\frac{340 \eta^2}{9} \right)
\label{eq.SPA1}
\eea
where $\delta\equiv (m_1-m_2)/(m_1+m_2)$, $\chi_s\equiv(\chi_1+\chi_2)/2$ and $\chi_a\equiv(\chi_1-\chi_2)/2$ with $\chi_i \equiv S_i/m_i^2$, $S_i$ being the spin angular momentum of the $i{\rm th}$ compact object. 
On the other hand, it is more efficient to treat the two
spin parameters with a single effective spin parameter $\chi_{\rm eff}$ because the two spins are
strongly correlated in an aligned-spin system \cite{Aji11,San10,Pur13,Pur16,Cho16}.
The effective spin is defined by
\be\label{eq.effective spin}
\chi_{\rm eff}={m_1\chi_1+m_2\chi_2 \over M}.
\ee
If we choose the same component spin values in the wave function,  the value of $\chi_{\rm eff}$ can be given as $\chi_{\rm eff}=\chi_1=\chi_2$.
We also use $\{M_c, \eta\}$ as our mass parameters rather than $\{m_1, m_2\}$ since those parameters are generally more efficient in the GW data analysis.

In equation (\ref{eq.Psi}), $\phi^{\rm Tidal}$ represents the tidal effect induced by the tidal deformation of neutron stars; thus, this term is not applicable to the BBH waveforms.
The tidal correction term first appears from 5PN order but its impact on the wave phase is comparable to the 3.5PN order phase term \cite{Fla08}.  
We here consider the tidal correction up to 6PN order as \cite{Fav14, Lac15}
\be \label{eq.tidal correction}
 \phi^{\rm Tidal} = - \bigg[ \frac{39 \tilde{\Lambda}}{2}v^{10} + \bigg( \frac{3115\tilde{\Lambda}}{64} - \frac{6595 \sqrt{1-4\eta} \delta \tilde{\Lambda}}{364} \bigg)v^{12} \bigg],
\ee
where $\tilde{\Lambda}$ and $\delta \tilde{\Lambda}$ are defined by \cite{Fav14, Lac15,Hin10}
\bea
\tilde{\Lambda}  &=& \frac{8}{13}[ (\rm 1+7\eta - 31\eta^2)( \Lambda_1 + \Lambda_2 )+ \sqrt{1 - 4\eta}(1 +9\eta -11\eta^2)(\Lambda_1 - \Lambda_2)], \nonumber\\
\delta \tilde{\Lambda} &=& \frac{1}{2} \bigg[  \sqrt{1-4\eta} \bigg(1 - \frac{13272\eta}{1319} + \frac{8944\eta^2}{1319} \bigg)( \Lambda_1 + \Lambda_2 ) \nonumber  \\
&+& \bigg( 1 - \frac{15910\eta}{1319} + \frac{32850\eta^2}{1319} + \frac{3380\eta^3}{1319} \bigg) ( \Lambda_1 - \Lambda_2 ) \bigg],
\eea
where $\rm \Lambda_i $ is the dimensionless tidal deformability. 
For simplicity, we only consider equal-mass BNSs, then $\tilde{\Lambda} = \Lambda_1 = \Lambda_2$
and  $\delta \tilde{\Lambda}$ term vanishes in equation (\ref{eq.tidal correction}).

\section{Parameter measurement error}    \label{sec.PE}
The waveforms emitted from compact binary systems can be modeled reasonably well, thus
the matched filter technique is likely to be used in the GW data analysis.
The matched filter output is given by using the inner product $\langle ... | ... \rangle$ as
\be \label{eq.match}
\langle x | h \rangle = 4 {\rm Re} \int_{f_{\rm low}}^{f_{\rm max}}  \frac{\tilde{x}(f)\tilde{h}^*(f)}{S_n(f)} df,
\ee
where the tilde denotes the Fourier transform of the time-domain
waveform, $x(t)$ is the detector data stream, $h(t)$ is the filter template,
$S_n(f)$ is the noise PSD for the detector,
$f_{\rm low}$ and $f_{\rm max}$ are the low and maximum frequency limits, respectively. 
Once a detection is made in the GW search pipeline,
a more detailed analysis is conducted in the parameter estimation pipeline 
to find the source properties, such as the distance, mass, spin, etc. \cite{Vei15}.
By means of  Bayesian statistics, the result of the parameter estimation analysis is given by the marginalized posterior probability density functions (PDFs)  for the parameters.
The posterior probability that the GW signal in the data stream $x$ is characterized by the parameters $\lambda$, can be expressed by 
\be \label{eq.posterior}
p(\lambda|x) \propto p(\lambda)L(x|\lambda), 
\ee
where $p(\lambda)$ is the prior probability, 
$L(s|\lambda)$ is the likelihood given by~\cite{Cut94,Fin92}
\be \label{eq.L0}
L(x|\lambda) \propto {\rm exp}\bigg[-\frac{1}2 \langle x-h(\lambda)|x-h(\lambda)\rangle\bigg],
\ee
where $h(\lambda)$ is a model waveform.
If the SNR is high enough, i.e., the detector noise is negligible with respect to the incident signal strength, 
the likelihood can be given by
\be \label{eq.L}
L\propto {\rm exp} \bigg[ -\frac{1}2\{\langle s|s \rangle+\langle h|h\rangle-2\langle s|h\rangle\} \bigg],
\ee
where $s$ is the GW signal.
With the assumption of flat priors in equation~(\ref{eq.posterior}), the Bayesian posterior PDF is determined only by the likelihood.

Since the parameter estimation algorithm explores the entire parameter space, the procedure usually takes a very long time.
On the other hand, the measurement errors of the parameters can be determined by using the Fisher matrix method which is a semi-analytic approach \cite{Val08,Cho15}.
Fisher matrix can estimate the $1\sigma$ statistical errors of parameters for a given signal with known true values.
Using the definition of the inner product given in equation~(\ref{eq.match}), the SNR ($\rho$) for the signal $s$ can be given by  \cite{All12} 
\be \label{eq.snr}
\rho=\sqrt{\langle s|s \rangle}.
\ee
If we choose a template whose waveform  is
similar to the signal waveform, we have $\rho^2 =\langle s|s \rangle \simeq \langle h|h \rangle$.
Then, we  can rewrite equation~(\ref{eq.L}) as~\cite{Cho13} 
\be \label{eq.L2}
L(\lambda) \propto {\rm exp} [-\rho^2\{1- \langle \hat{s}|\hat{h}(\lambda)\rangle\}],
\ee
where $\hat{h}$ denotes the normalized waveform defined by $\hat{h}= h/ \rho$.
Generally, the likelihood has the  form of a Gaussian distribution in the high SNR limit~\cite{Cut94,Fin92}.

If one describes a Gaussian function centred around $x_0$  as
\be
p(x)\propto e^{-\Gamma \Delta x^2/2},
\ee
where $\Delta x=x-x_0$,
then the variance $\sigma^2$ corresponds to $\Gamma^{-1}$,
and $\Gamma$ can be calculated by
\be
\Gamma=-{d^2 \ln p(x) \over d^2 x}\bigg|_{x=x_0}.
\ee
Similarly, for a N-dimensional case, $\Gamma_{ij}$ is a $N\times N$ matrix and the Gaussian function can be expressed as
\be
 p(x_i)=e^{-\Gamma_{ij} \Delta x_i \Delta x_j/2}.
 \ee
In this case, the matrix $\Gamma_{ij}$ is determined by
\be \label{eq.GammaNd}
\Gamma_{ij}=-{\partial^2 \ln p(x) \over \partial x_i \partial x_j}\bigg|_{x=x_{0}}.
\ee

If we assume flat priors in equation~(\ref{eq.posterior}), we have $p(\lambda|x) \propto L(x|\lambda)$.
By substituting $L(x|\lambda)$ for $p(x)$ in equation~(\ref{eq.GammaNd}), and using equation~(\ref{eq.L2}) and the relation $\hat{s} = \hat{h}_0$, we can  get
\be \label{eq.Gamma}
\Gamma_{ij}=- {\partial^2  \ln L \over \partial \lambda_i \partial \lambda_j}\bigg|_{\lambda=\lambda_0} 
=-\rho^2 {\partial^2   \langle \hat{h}_0|\hat{h}(\lambda)\rangle \over \partial \lambda_i \partial \lambda_j}\bigg|_{\lambda=\lambda_0}.
\ee
Using this equation and the relation $h=\rho \hat{h}_0$, the matrix $\Gamma_{ij}$ can be obtained directly from the wave function $h(\lambda)$ as follows \cite{Val08,Jar94}
\be\label{eq.generalFM}
\Gamma_{ij} = \bigg \langle {\partial h \over \partial \lambda_i} \bigg| {\partial h \over \partial \lambda_j} \bigg \rangle\bigg|_{\lambda=\lambda_0}.
\ee
This is the familiar expression of the standard Fisher matrix introduced in \cite{Cut94, Poi95}.
The above procedure implies that the Fisher matrix can be obtained by taking only the leading-order terms of the Taylor expansion in all parameters of the Likelihood about the maximum likelihood point ($\lambda=\lambda_0$)
and this approximation is only valid in the high SNR limit.
The inverse of Fisher matrix represents the covariance matrix ($\Sigma_{ij} = (\Gamma^{-1})_{ij} $) of the parameter errors, 
and the measurement error $\sigma_i$ (standard deviation of a normal distribution) of each parameter and the correlation coefficient (${\cal C}_{ij}$) between two parameters are given by
\be \label{eq.sigma}
\sigma_i=\sqrt{\Sigma_{ii}},  \ \ \ {\cal C}_{ij}={\Sigma_{ij} \over \sqrt{\Sigma_{ii} \Sigma_{jj}}}.
\ee
From equations (\ref{eq.Gamma}) and (\ref{eq.sigma}), one can see that the measurement error is inversely proportional to the SNR
for a given signal $\hat{s}$.

In Bayesian parameter estimation simulation, any form of  prior distribution can be imposed on the analysis.
However, in the Fisher matrix formalism, only the Gaussian prior function can be applied analytically.
If the parameter $\lambda_i$ has a Gaussian prior distribution centered at the true value $\lambda_0$ with variance $P_{\lambda}^2$,
the (prior-incorporated) covariance matrix can be determined simply by adding the component $ \Gamma_{ii}^0=P_{\lambda}^{-2}$ as \cite{Cut94, Poi95}
\be  \label{eq.error-correl}
\Sigma^P_{ij} = (\Gamma_{ij} + \Gamma_{ii}^0)^{-1}, \label{eq.covariant_matrix_with_prior}
\ee
where $ \Gamma_{ij}$ is the original Fisher matrix in which no prior information is imposed.

One element of the Fisher matrix can be determined independently of the computation of the other elements. 
Although an N-dimensional Fisher matrix rises to N+n dimension by adding n parameters to the original Fisher matrix,
the original elements remain unchanged.
However, all elements of the original N-dimensional covariance matrix can be changed by the addition of the n parameters. 
Typically, the measurement errors of the intrinsic parameters can significantly increase by adding another intrinsic parameter
because the intrinsic parameters are usually strongly correlated with each other.
In contrast, adding extrinsic parameters to the Fisher matrix does not affect the result of the intrinsic parameters
due to the weak correlation between the extrinsic and intrinsic parameters.

One should note that the error calculated by $\sigma_i=\sqrt{(\Gamma_{ii})^{-1}}$ is different from the error calculated by $\sigma_i=\sqrt{(\Gamma^{-1})_{ii}})$.
The former represents the error of the parameter $\lambda_i$ obtained by using only $i$th diagonal element of the N-dimensional Fisher matrix,
in this case, the true values of the other parameters are assumed to be exactly known.
Meanwhile, the latter  represents the error of $\lambda_i$ obtained by using the inversion of the N-dimensional Fisher matrix as described in equation (\ref{eq.error-correl}),
and in that case, the error $\sigma_{\lambda_i}$ corresponds to the marginalized one-dimensional error that incorporates the correlations with the other parameters.
All errors given in this work are of the latter case.


\section{Result :  comparison between aLIGO and ET} \label{sec.result}
\subsection{Binary black holes}
We apply the TaylorF2 wave function to the Fisher matrix formalism given in equation~(\ref{eq.generalFM}) to calculate the measurement errors.
For the BBH system, the Fisher matrix is given by a $5\times 5$ matrix with the three intrinsic parameters $\{ M_c, \eta, \chi_{\rm eff}\}$ and $\{t_c, \phi_c \}$. 
The two parameters $t_c$ and $\phi_c$ are arbitrary constants, so the choice of their true values does not affect the errors of the intrinsic parameters.
However, since those parameters are strongly correlated with the intrinsic parameters, 
omitting them significantly affects the errors of the intrinsic parameters (see, table \ref{tab.5dvs6d}).
Although $t_c$ and $\phi_c$ are physically uninteresting parameters,
they must be included in the construction of the Fisher matrix.
In this work, thus, we consider $t_c$ and $\phi_c$ together with the intrinsic parameters in the Fisher matrix but
present only the results for the intrinsic parameters.
We consider an L-shaped single-detector configuration and use the aLIGO and the ET detector sensitivity curves given in figure~\ref{fig.psd}.
According to the shape of the sensitivity curves, the low-frequency limit is given as $f_{\rm low}=10 \ (1)$ Hz for aLIGO (ET).
The maximum-frequency limit is given as $f_{\rm max}=f_{\rm ISCO}$ for both the cases, where the ISCO frequency is given by
\be
f_{\rm ISCO}={1 \over  6^{3/2}\pi M}.
\ee
Although the masses of the BBHs detected by aLIGO and Advance Virgo are broadly distributed between $10 \msun$ and $160 \msun$ \cite{GWTC-1,GWTC-2},
we only consider the low-mass BBHs with $M\leq 16 \msun$ due to the validity of the TaylorF2 model as discussed above.
We also consider the component masses of $m_1, m_2 \geq3\msun$ and the effective spin of $-0.9\leq \chi_{\rm eff} \leq 0.9$.
We assume the effective distance to be $D_{\rm eff}=400$ Mpc for all BBH sources.
For the errors of the chirp mass in our result, we use the dimensionless relative error $\sigma_{M_c}/M_c$, while we use the absolute error $\sigma_{\lambda}$ for the other parameters.
Finally, for completeness, it is worth mentioning that using the same waveform, parameters, and PSD as  \cite{Poi95}  we are able to reproduce their result with greater than $99 \%$ accuracy.

\begin{figure}[t]
\begin{center}
\includegraphics[width=\columnwidth]{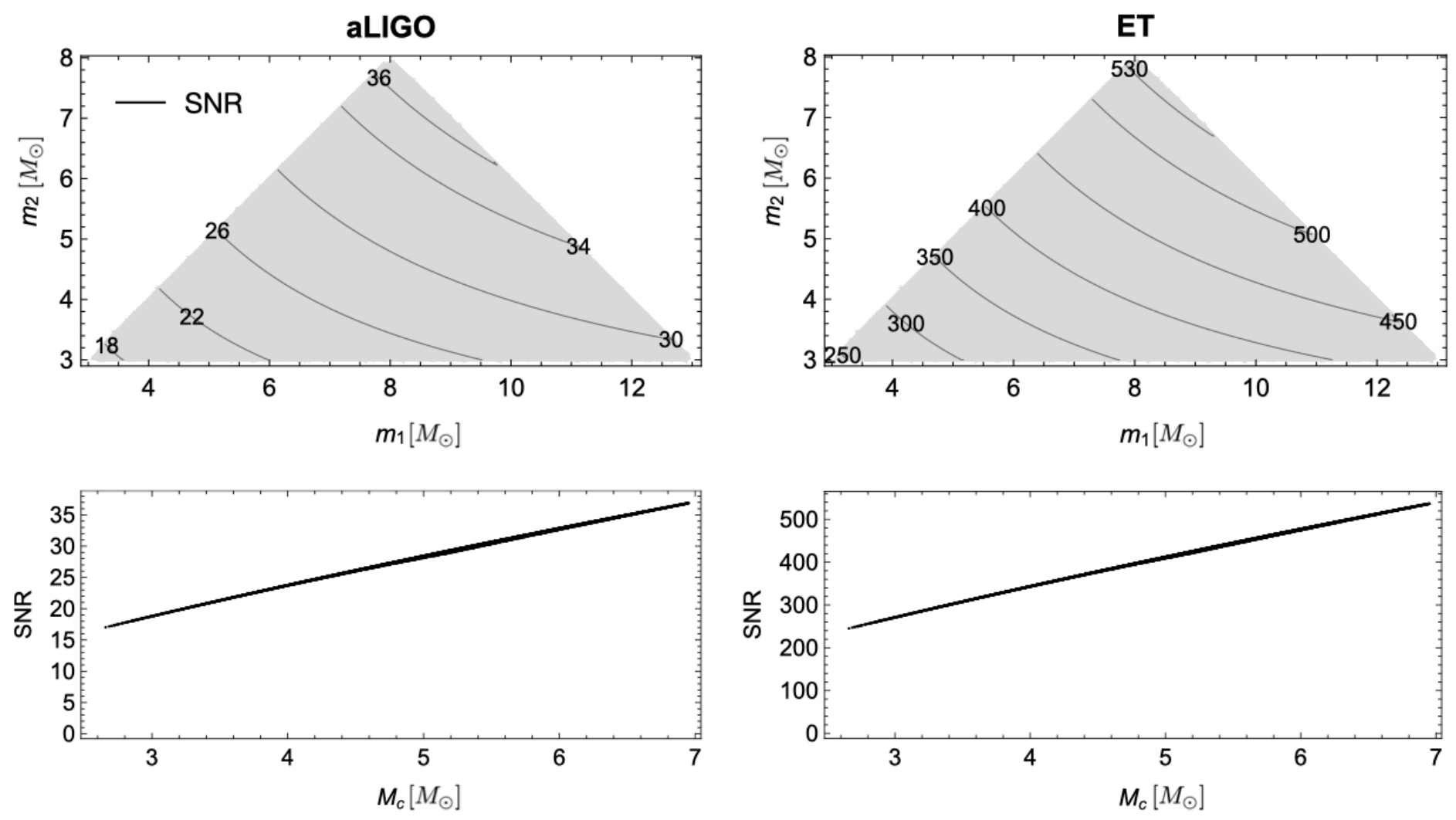}
\caption{\label{fig.snr} The SNRs calculated with $10^4$ Monte-Carlo BBH samples. The samples are randomly generated in the range of 
$m_2, m_2 \geq 3\msun, M\leq 16\msun, -0.9 \leq \chi_{\rm eff} \leq 0.9$
assuming $D_{\rm eff}=400{\rm Mpc}$.
The lower panels show the strong dependence of SNR on $M_c$ ($\rho \propto M_c^{5/6})$ and negligible dependence on $\eta$ and $\chi_{\rm eff}$. For all binaries, we have ${\rm SNR_{ET}/SNR_{aLIGO}} \sim 14$.}
\end{center}
\end{figure}

We generate $10^4$ Monte-Carlo BBH samples, which are randomly distributed in the $m_1$--$m_2$--$\chi_{\rm eff}$ parameter space in the range of $m_1, m_2 \geq3\msun (m_1 \geq m_2), M \leq 16\msun,$ and $-0.9\leq \chi_{\rm eff} \leq 0.9$.
Using equation (\ref{eq.snr}), we calculate the SNRs for all binaries for the aLIGO and the ET cases, respectively.
In figure \ref{fig.snr}, the upper panel shows the SNRs in the $m_1$--$m_2$ plane.
Note that in this result, we use all of the $10^4$ samples, so the SNR dependence on $\chi_{\rm eff}$ is projected onto the $m_1$--$m_2$ plane.
We have the SNRs  of $17 \lesssim \rho \lesssim 37$  for aLIGO and  $240 \lesssim \rho \lesssim 540$ for ET, 
and we find that ET can achieve $\sim 14$ times larger SNRs than aLIGO for all binaries. 
Most of the signal strength is determined by its amplitude, so the SNR is only a function of $M_c$ (i.e., $\rho \propto M_c^{5/6}$) for a given $D_{\rm eff}$.
This relation is clearly described in the lower panel.
In this panel, we also use all of the $10^4$ samples, so the small width of the curves indicates a negligible dependence of $\eta$ and $\chi_{\rm eff}$.

\begin{figure}[t]
           \includegraphics[width=\columnwidth]{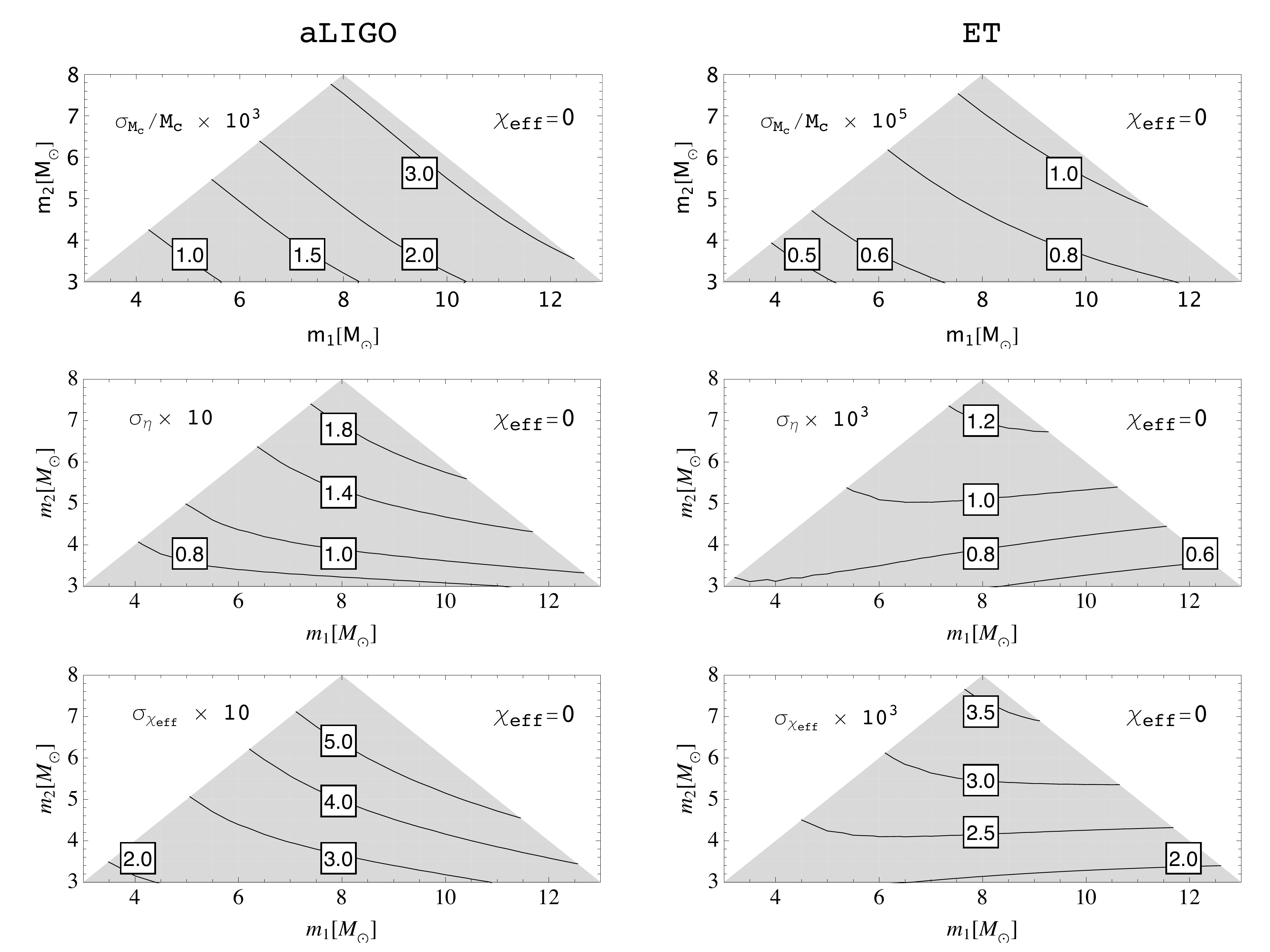}  
                  \caption{\label{fig.BBH-error} Measurement errors of the chirp mass (top), the symmetric mass ratio (middle), and the effective spin (bottom) for the binaries with $\chi_{\rm eff}=0$. We assmue $D_{\rm eff}=400$ Mpc. }
\end{figure}

Figure \ref{fig.BBH-error} shows the measurement errors of the chirp mass, the symmetric mass ratio, and the effective spin obtained with aLIGO and ET for the sources with $\chi_{\rm eff}=0$. 
Among the three parameters, the most accurately measured one is the chirp mass, of which the measurement errors are  $\sigma_{M_c}/M_c \sim {\cal O}(10^{-3})$ for aLIGO and $\sim {\cal O}(10^{-5})$ for ET. The measurement errors of  $\eta$ and $\chi_{\rm eff}$ are $\sigma_{\eta, \chi_{\rm eff}} \sim {\cal O}(10^{-1})$ for aLIGO and $\sim {\cal O}(10^{-3})$ for ET, but the accuracy for $\eta$ is 2--4 times better.
One can see that the parameter measurement between the two detectors is greatly improved. 
Overall, the measurement errors for ET can be lower than 1\% of those for aLIGO.
The trends of the contours for aLIGO and ET  are similar for $\sigma_{M_c}$ but quite different for $\sigma_{\eta}$ and $\sigma_{\chi_{\rm eff}}$.
For aLIGO, $\sigma_{\eta}$ and $\sigma_{\chi_{\rm eff}}$ have the smallest values when the binary mass is smallest, i.e. $(3\msun, 3\msun)$,
but for ET, they have the smallest values when the component masses are most asymmetric even though the binary is the heaviest in our mass range, i.e. $(13\msun, 3\msun)$.
On the other hand, the results of the symmetric binaries ($\eta \sim 1$) show that 
the measurement error of the intrinsic parameter increases as $M$ increases even though the signal strength becomes stronger.
This can be understood by the fact that 
the GW signal of a less massive system (for a given value of $\eta$) contains more wave cycles as the frequency sweeps through the detector bandwidth,
and the phasing of the signal plays the largest role in measuring the mass and the spin parameters.
(Clearly, this is only true for the intrinsic parameters, which affect the wave phasing.
The measurement of the extrinsic parameters mainly depends on the SNR because they are integrated into the wave amplitude.)

\begin{table}
\begin{center}
\caption{\label{tab.intrinsic-errors}{Parameter measurement errors and SNRs for BBHs with aLIGO and ET. The effective distance is 400 Mpc. 
The ratios are defined by $ {{\cal R}_{\lambda}} = \sigma_{\lambda,\rm{ET}} /  \sigma_{\lambda,\rm{aLIGO}}$ and  ${\cal R}_{\rm SNR}  = {\rm SNR_{ET}/SNR_{aLIGO}}$.}}
\begin{indented}
\item[]\begin{tabular}{ ccc  ccc   c}
\multicolumn{6}{c}{aLIGO}\\
\br
$m_1 [\msun]$    &$m_2[\msun]$& $ \chi_{\rm eff} $&$\sigma_{M_c }/M_c {\times} {10^3}$&$\sigma_{\eta}$&$\sigma_{\chi_{\rm eff}}$ &  SNR \\
\mr
  3.0&3.0   &0.0  & 0.61 &  0.061  &   0.175  &16.8 \\
    5.0& 5.0  &0.0   & 1.29 &  0.100    &   0.295   & 25.5 \\
8.0& 8.0   &0.0 &3.20 &  0.208   &   0.622   & 37.0 \\
    13.0&3.0  &0.0   &2.83 &  0.086  &   0.353  &28.9 \\
\br
\\
\multicolumn{6}{c}{ET}\\
\br
$m_1 [\msun]$    &$m_2[\msun]$& $ \chi_{\rm eff} $&$\sigma_{M_c }/M_c {\times} {10^5}$&$\sigma_{\eta}{\times} {10^3}$&$\sigma_{\chi_{\rm eff}}{\times} {10^2}$ &  SNR \\
\mr
   3.0&3.0  &0.0   & 0.38 & 0.78  &   0.203  &241.5 \\
    5.0& 5.0  &0.0   & 0.63 &  0.96    &   0.265   & 368.2 \\
8.0& 8.0   &0.0 &1.07&  1.26  &   0.361 & 537.5 \\
    13.0&3.0  &0.0   &0.85 &  0.46  &   0.177  &419.6 \\\br
\\
\multicolumn{6}{c}{Comparison}\\
\br
$m_1 [\msun]$    &$m_2[\msun]$& $ \chi_{\rm eff} $&${\cal R}_{M_c}{\times} {10^2}$&${\cal R}_{\eta}{\times} {10^2}$&${\cal R}_{\chi_{\rm eff}}{\times} {10^2}$ &${\cal R}_{\rm{SNR}}$\\\mr
 3.0&3.0  &0.0   & 0.63 &  1.27  &   1.16  &14.3 \\
    5.0& 5.0   &0.0  & 0.49&  0.96   &   0.90   & 14.4 \\
8.0& 8.0 &0.0   &0.33 &  0.61   &   0.58  & 14.5 \\
    13.0&3.0  &0.0   &0.30 &  0.54  &   0.50  &14.5 \\  \br

\end{tabular} 
\end{indented}
\end{center}
\end{table}

\begin{figure}[t]
\begin{center}
           \includegraphics[width=\columnwidth]{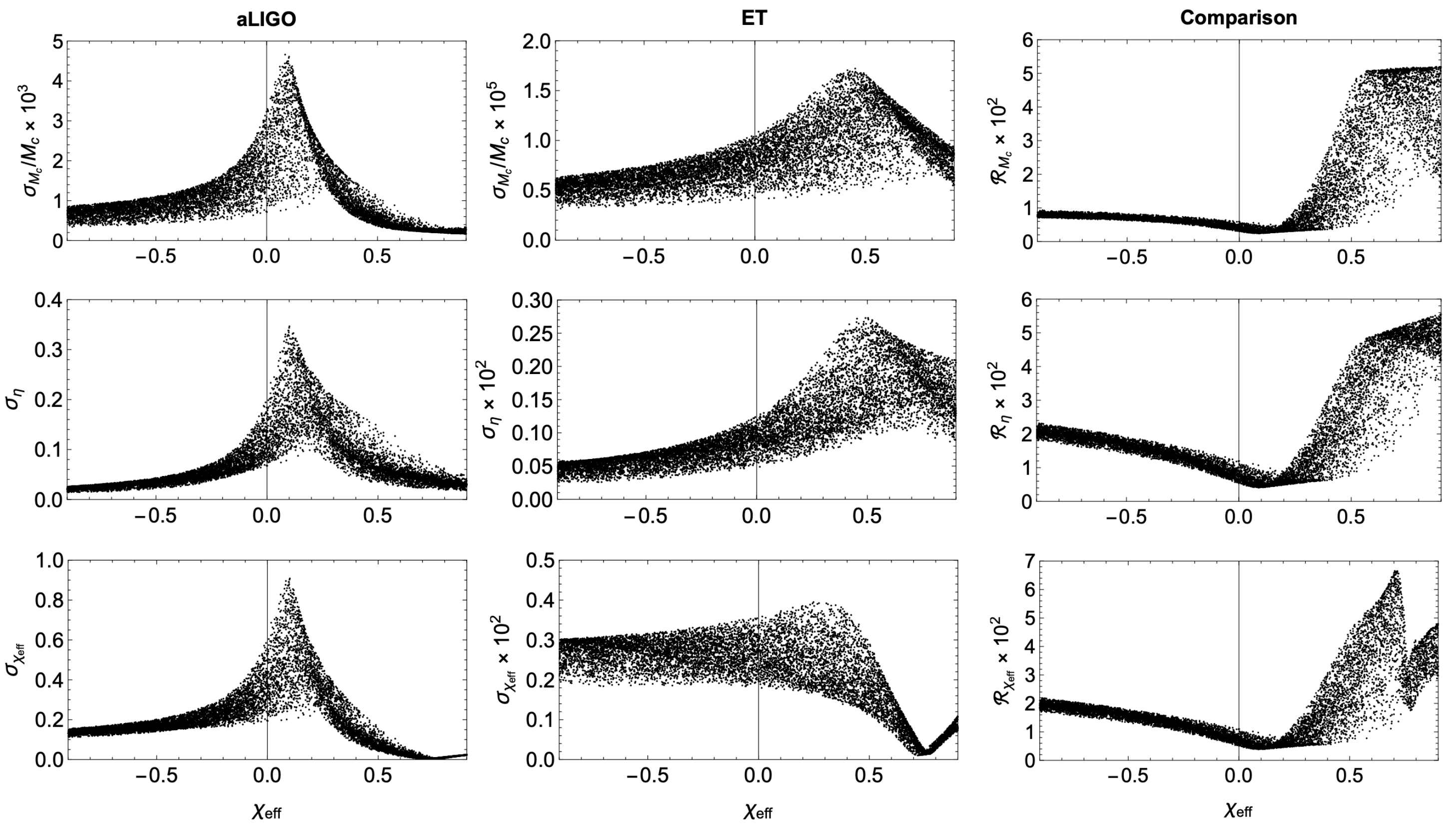}  
                  \caption{\label{fig.MonteCarlo-errors} Measurement errors  obtained from the $10^4$ Monte-Carlo simulations. For comparison between aLIGO and ET, we define the error ratio as $ {{\cal R}_{\lambda}} = \sigma_{\lambda,\rm{ET}} /  \sigma_{\lambda,\rm{aLIGO}}  $}
\end{center}
\end{figure}

\begin{figure}[t]
\begin{center}
           \includegraphics[width=\columnwidth]{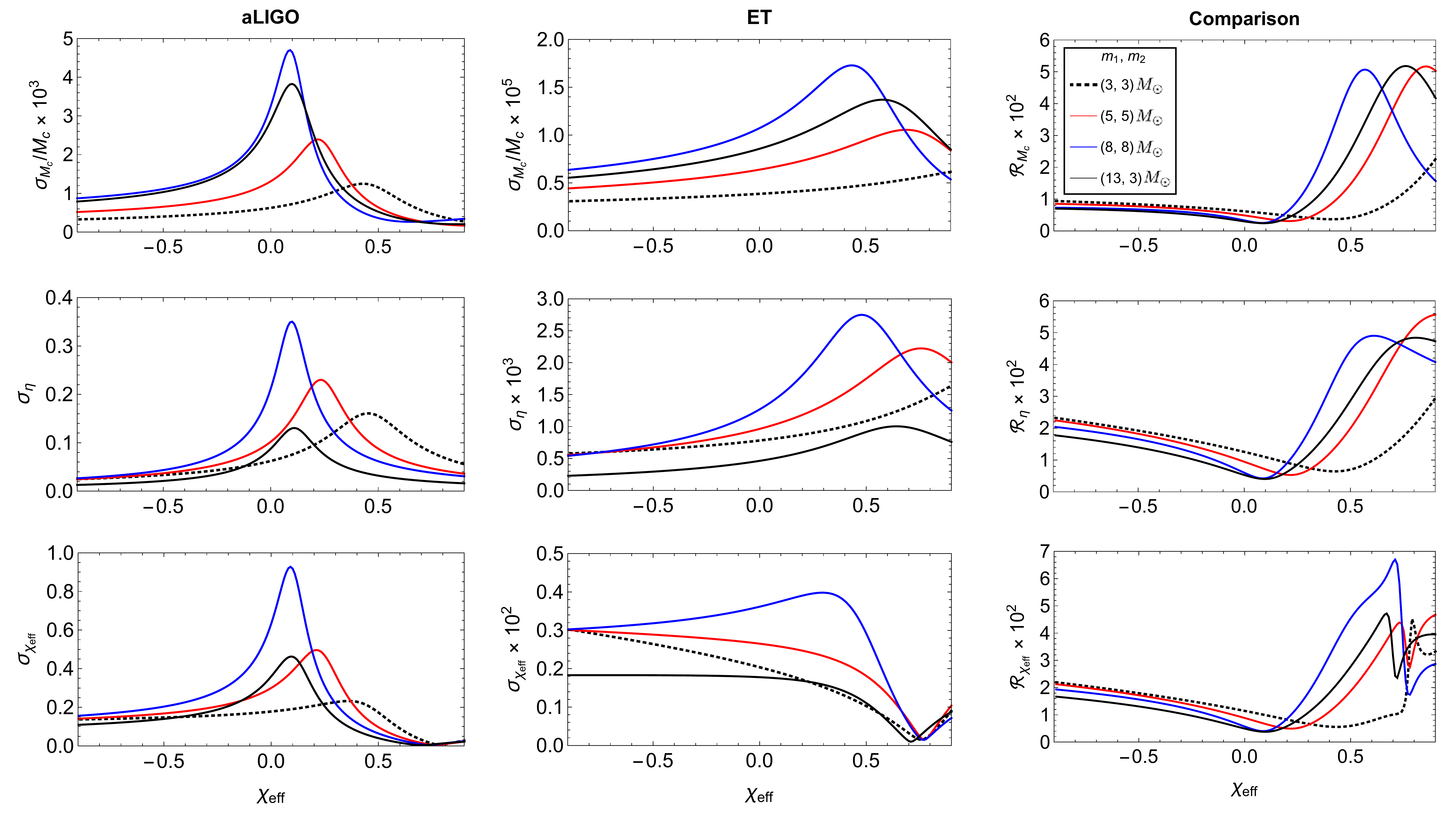}  
                  \caption{\label{fig.1d-error-comparison} One-dimensional error curves given as a function of $\chi_{\rm eff}$ for the selected binaries with different masses.}
\end{center}
\end{figure}

\begin{figure}[t]
           \includegraphics[width=\columnwidth]{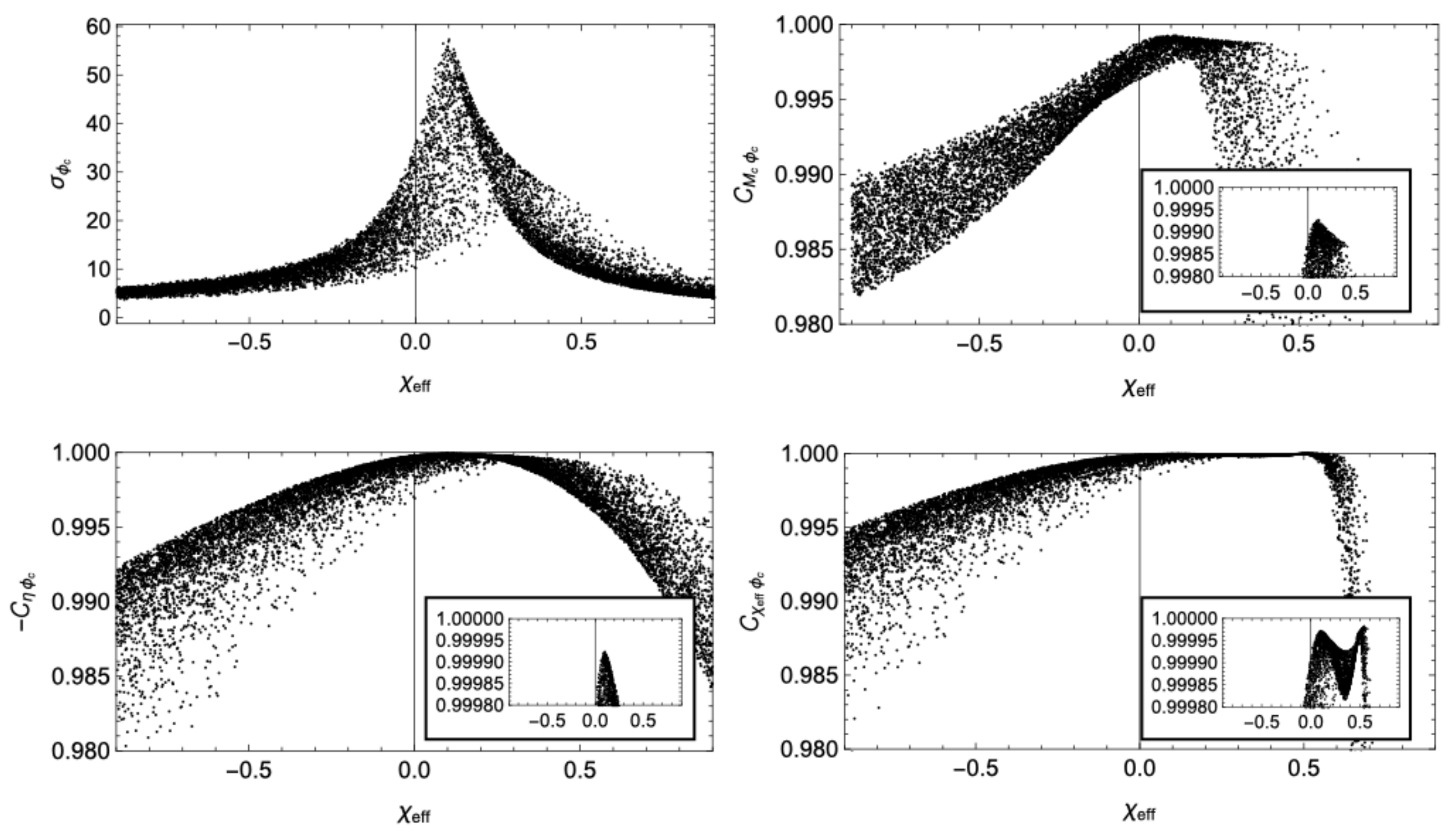}  
                  \caption{\label{fig.MonteCarlo-correlation} Measurement error of $\phi_c$ (upper left) and the correlations with the intrinsic parameters for aLIGO. All correlations have very high values at the same region around $\chi_{\rm ef} = 0.1$ (see, the inset). Note that the correlation ${\cal C}_{\eta \phi_c}$ is negative.}
\end{figure}

To see the spin dependence of the error, we display the errors as a function of $\chi_{\rm eff}$ in figure \ref{fig.MonteCarlo-errors}.
The distribution of errors differs significantly between aLIGO and ET, showing no symmetry about $\chi_{\rm eff}=0$ in both cases.
The result shows sharp peaks at $\chi_{\rm eff} \sim 0.1$ for aLIGO but smooth peaks around $\chi_{\rm eff} = 0.4$ for ET. 
At $\chi_{\rm eff} =-0.9$, the error ratios are ${\cal R}_{M_c} \leq 1 \%$ and ${\cal R}_{\eta}, {\cal R}_{\chi_{\rm eff}} \leq 2.5 \%$, 
and they slowly decrease with $\chi_{\rm eff}$ and from $\chi_{\rm eff} \sim 0.2$, they are widely distributed up to $5 - 7 \%$.
Although not shown here, for all of the samples we have $\sigma_{t_c}<0.02 \ (0.0035)$ s for aLIGO (ET).
For easy understanding of the distribution of the errors, 
we select four BBH sources with different masses and display the one-dimensional error curves in figure \ref{fig.1d-error-comparison}.
We find that the curves coincide with the error distributions.
We also find that the upper part of the error distribution can be obtained from the symmetric and higher-mass binaries ($m_1, m_2\sim 8\msun$), except for the high-spin region.
In table \ref{tab.intrinsic-errors}, we present exact values of the SNR, $\sigma_{\lambda}$,  and ${\cal R}_{\lambda}$ for the selected binaries with $\chi_{\rm eff}=0$.
On the other hand, in the case of aLIGO, we figure out that the sharp peaks are caused by the 
strong correlations between the intrinsic parameters and $\phi_c$.
In figure \ref{fig.MonteCarlo-correlation}, one can see that for aLIGO
the errors of $\phi_c$ also have a sharp peak at $\chi_{\rm ef} \sim 0.1$,
and all correlations with the intrinsic parameters can have very high values at the same region around $\chi_{\rm ef} = 0.1$.
Consequently, the binary samples in the high-correlation area constitute these sharp peaks.
Note that none of the high-correlation samples encounters the matrix inversion problem in our Fisher matrix computations (see \ref{ap.inversion}).
For ET, the errors $\sigma_{\phi_c}$ for all samples are less than 0.4.

\subsection{Effectt of prior}

\begin{figure}[t]
           \includegraphics[width=\columnwidth]{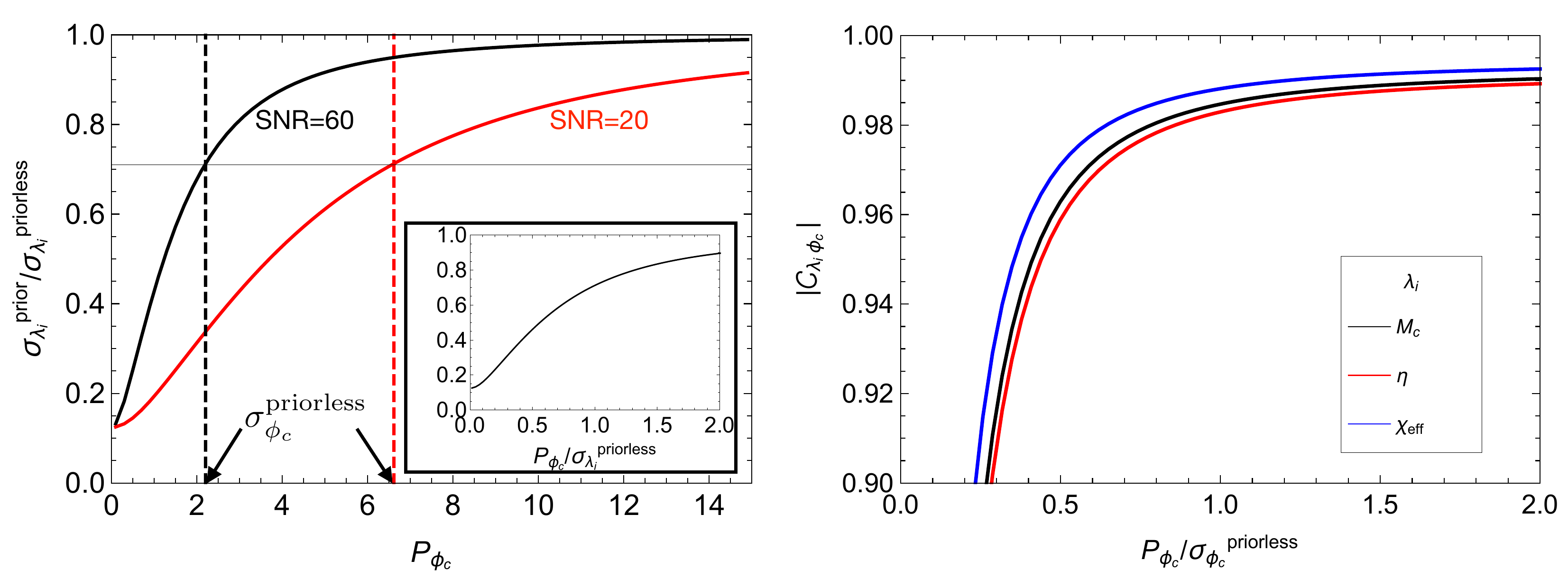}  
                  \caption{\label{fig.prior-dependence} Left: Ratio  between the prior-imposed measurement error ($\sigma^{\rm prior}$) and the priorless error ($\sigma^{\rm priorless}$) for the intrinsic parameters $\lambda_i=(M_c, \eta, \chi_{\rm eff})$. $P_{\phi_c}$ represents the standard deviation of the Gaussian prior function.
                  The virtical dotted lines indicate $\sigma_{\phi_c}^{\rm priorless}$ for the SNRs of 60 (black) and 20 (red). If we rescale the x-axis to $P_{\phi_c}/\sigma^{\rm priorless}_{\Phi_c}$, the two curves are exactly the same and independent of the SNR (see, the inset). Right: Correlations between $\lambda_i$ and $\phi_c$ given as a function of $P_{\phi_c}/\sigma^{\rm priorless}_{\Phi_c}$.   Here, we use a binary with the true values of $(m_1, m_2, \chi_{\rm eff})=(6\msun, 3\msun, -0.5)$.
}
\end{figure}

As discussed in section \ref{sec.PE}, the measurement error can be reduced by imposing the prior information
if the prior distribution is narrow enough.
To evaluate the impact of prior,  using equation (\ref{eq.covariant_matrix_with_prior}), 
we calculate the prior-imposed measurement error ($\sigma^{\rm prior}$) varying the prior variance $P^2$ 
and compare that with the original priorless error ($\sigma^{\rm priorless}$).
In the left panel of figure \ref{fig.prior-dependence}, we display the ratio $\sigma^{\rm prior}/\sigma^{\rm priorless}$  as a function of $P_{\phi_c}$ for the intrinsic parameters
assuming a binary with the true values of $(m_1, m_2, \chi_{\rm eff})=(6\msun, 3\msun, -0.5)$.
This result describes how the measurement error can be reduced as the prior variance decreases.
In equation (\ref{eq.covariant_matrix_with_prior}), the original Fisher matrix depends on the SNR, while the prior variance ($P^2$) is independent of the SNR.
Then, it can be shown that the prior contribution is stronger for lower SNRs, which is described in the left panel of figure \ref{fig.prior-dependence}.
In this panel, the upper and lower curves are obtained by assuming $\rho=60$ and 20, respectively.
If we rescale the x-axis to $P_{\phi_c}/\sigma^{\rm priorless}_{\Phi_c}$, the two curves are exactly the same (see, the inset).
Therefore, the prior contribution can be generalized independently of the SNR 
by defining the quantity $\sigma^{\rm prior}_{M_c}/\sigma^{\rm priorless}_{M_c}$ as a function of $P_{\phi_c}/\sigma^{\rm priorless}_{\Phi_c}$.
For example, if $P_{\phi_c}/\sigma^{\rm priorless}_{\Phi_c} \sim 1$, then $\sigma^{\rm prior}_{\lambda} \sim 70 \%$ of $\sigma^{\rm priorless}_{\lambda}$.
As seen in equation (\ref{eq.covariant_matrix_with_prior}), adding the component $ \Gamma_{ii}^0=P_{\theta}^{-2}$ to the original Fisher matrix changes every element of the covariance matrix, so all correlations change.
The right panel of figure \ref{fig.prior-dependence} shows the correltions between the intrinsic parameters and $\phi_c$ as a function of $P_{\phi_c}/\sigma^{\rm priorless}_{\Phi_c}$.
This result shows that the strong correlations rapidly weaken as the prior variance decreases.

\begin{figure}[t]
           \includegraphics[width=\columnwidth]{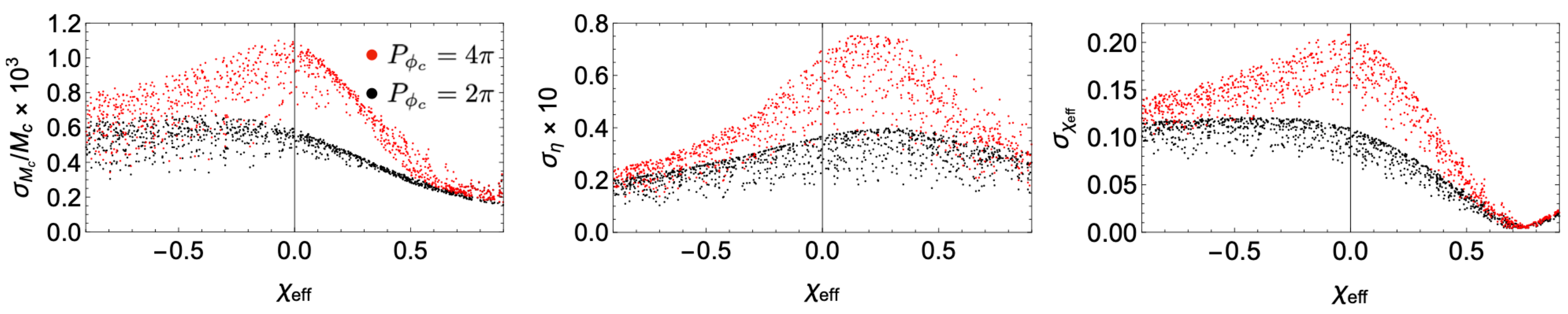}  
                  \caption{\label{fig.MonteCarlo-errors-prior} Measurement errors obtained by imposing the Gaussian prior on $\phi_c$ ($P_{\phi_c}=2\pi, 4\pi$) using $10^3$ Monte-Carlo samples.}
\end{figure}

One can see in figure \ref{fig.MonteCarlo-correlation} that for aLIGO a fairly large number of samples can produce the errors $\sigma_{\phi_c}$ much larger than $2\pi$,
which is a physically reasonable range for $\phi_c$.
We apply the Gaussian prior on $\phi_c$ ($P_{\phi_c}=2\pi, 4\pi$) to our analysis and recalculate the measurement errors of the intrinsic parameters using $10^3$ Monte-Carlo samples.
The result is given in figure \ref{fig.MonteCarlo-errors-prior}.
For the case of $P_{\phi_c}=4\pi$,  the errors are greatly reduced, and the sharp peaks are relieved, compared to the result in figure \ref{fig.MonteCarlo-errors}.
The errors for $P_{\phi_c}=2\pi$ are more reduced by about a factor of 2 compared to those for $P_{\phi_c}=4\pi$, as expected.
We also perform the same analysis for ET 
but the result is indistinguishable from that in figure \ref{fig.MonteCarlo-errors}.
This is because the errors $\sigma_{\phi_c}$ for ET are much smaller than $2\pi$ and the SNRs are high enough.
In this case, the posteriors are almost unaffected by the prior $P_{\phi_c} \sim 2\pi$ as described in figure \ref{fig.prior-dependence}.
Concrete examples for the effect of priors are also given in table 5 of \cite{Nielsen_2013}.


\subsection{BNSs}
For a BNS system, our Fisher matrix is given by a $6\times 6$ matrix with the parameters $\{ M_c, \eta, \chi_{\rm eff}, \tilde{\Lambda}, t_c, \phi_c \}$ assuming a fixed effective distance. 
The choice of $f_{\rm low}$ and $f_{\rm max}$ is the same as in the case of BBHs
but the effective distance is assumed to be  40 Mpc in order to obtain sufficiently high SNRs.
We consider three equal-mass binaries ($1\msun, 1\msun$), ($1.4\msun, 1.4\msun$), and ($2\msun, 2\msun$).
We choose the true value of the tidal parameter $\tilde{\Lambda}$ using the soft APR4 equation-of-state (EOS) model and the stiffer EOS model MPA1 given in \cite{PhysRevD.79.124032}.

\begin{table}
\begin{center}
\caption{\label{tab.NS-error}{Parameter measurement errors and SNRs for BNSs with aLIGO and ET. The effective distance is 40 Mpc. 
The ratios are defined by $ {{\cal R}_{\lambda}} = \sigma_{\lambda,\rm{ET}} /  \sigma_{\lambda,\rm{aLIGO}}$ and  ${\cal R}_{\rm SNR}  = {\rm SNR_{ET}/SNR_{aLIGO}}$.
For the errors of $M_c, \eta,$ and $\chi_{\rm eff}$ we only present the result for the MPA1 EOS model but the difference from those for ARP4 is within $1\%$.
For convenience, we list the result for $\tilde{\Lambda}$ separately, the error of $\tilde{\Lambda}$ depends on both the mass and the EOS.
}}
\begin{indented}
\item[]\begin{tabular}{ccc cc  c}
\multicolumn{6}{c}{aLIGO}\\
\br
$m_i [\msun]$    & $\chi_{\rm eff}$  &  $\sigma_{M_c }/M_c {\times} {10^5}$  &  $\sigma_{\eta}{\times} {10^3}$ & $\sigma_{\chi_{\rm eff}}{\times} {10^2}$   &  SNR \\
\mr
   2.0& 0.0   & 5.44&7.60  &  2.04     &120.0 \\
    1.4& 0.0      &3.64 &6.26 &1.62   & 89.2\\
  1.0& 0.0    & 2.58 &5.42  &  1.34  &67.4 \\
\br
\\
\multicolumn{6}{c}{ET}\\
\br
$m_i [\msun]$    & $\chi_{\rm eff}$  &  $\sigma_{M_c }/M_c {\times} {10^7}$  &  $\sigma_{\eta}{\times} {10^5}$  & $\sigma_{\chi_{\rm eff}}{\times} {10^4}$ &   SNR \\
\mr
    2.0&  0.0   & 3.19&9.33& 2.19 & 1724.8 \\
      1.4& 0.0      & 2.29 &8.29 &1.82   &1281.6 \\
   1.0& 0.0    & 1.68&7.50&  1.53 &968.3 \\
\br
\\
\multicolumn{6}{c}{ Comparison}\\
\br
$m_i [\msun]$    &  $\chi_{\rm eff}$  &${\cal R}_{M_c}{\times} {10^2}$&${\cal R}_{\eta}{\times} {10^2}$&  ${\cal R}_{\chi_{\rm eff}}{\times} {10^2}$ &${\cal R}_{\rm{SNR}}$\\
\mr
   2.0&  0.0   & 0.587&1.23&   1.07   &14.4\\
    1.4& 0.0     & 0.627&1.32&     1.13  &14.4\\
    1.0& 0.0      & 0.653&1.39 &  1.14&  14.4\\
\br
\\
\multicolumn{6}{c}{Result for $\tilde{\Lambda}$}\\
\br
$m_i [\msun]$    & EOS&  $\tilde{\Lambda} $ &   $\sigma_{\tilde{\Lambda}}\ (\sigma_{\tilde{\Lambda}}/\tilde{\Lambda})$ [aLIGO]  & $\sigma_{\tilde{\Lambda}}\ (\sigma_{\tilde{\Lambda}}/\tilde{\Lambda})$ [ET]  &${\cal R}_{\tilde{\Lambda}}{\times} {10^2}$\\
\mr
   2.0& MPA1&48  & 35.8 (0.746)&2.09 (0.0435) &  5.84     \\
     2.0& APR4& 15   & 34.1 (2.273)&2.07 (0.138) &  6.08     \\
    1.4& MPA1&513      &75.9 (0.148)&3.89 (0.00758) &5.13 \\
      1.4&  APR4&250       &63.5 (0.254)&3.77 (0.0151) & 5.93\\
  1.0& MPA1&3150    & 218.3 (0.0693)&7.94 (0.00252) &3.64 \\
    1.0& APR4& 1734    & 154.9 (0.0893)&7.27  (0.00419)& 4.69\\
\br
\end{tabular}
\end{indented}
\end{center}
\end{table}

In table \ref{tab.NS-error}, we summarize the SNRs and the measurement errors of the intrinsic parameters for BNSs with aLIGO and ET
and show a comparison between aLIGO and ET.
We have the SNRs  of $70 \lesssim \rho \lesssim 120$ for aLIGO and  $970 \lesssim \rho \lesssim1720$ for ET.
As in the case of BBHs, we find the SNR ratio of ${\cal R}_{\rm{SNR}}\sim 14$.
The measurement error of the chirp mass is  $\sigma_{M_c}/M_c \sim {\cal O}(10^{-5})$ for aLIGO and  $\sim {\cal O}(10^{-7})$ for ET. 
The measurement errors of  the symmetric mass ratio and the effective spin are $\sigma_{\eta}\sim {\cal O}(10^{-3}), \ \sigma_{\chi_{\rm eff}}\sim {\cal O}(10^{-2})$ for aLIGO and $\sigma_{\eta}\sim  {\cal O}(10^{-5}), \ \sigma_{\chi_{\rm eff}}\sim {\cal O}(10^{-4})$ for ET.
As in the case of BBHs, as $M$ increases the measurement accuracy for $M_c, \eta,$ and $\chi_{\rm eff}$ decreases independently of the increasing SNR.
Overall, the error ratios are ${\cal R}_{M_c} \leq 0.7 \%, {\cal R}_{\eta} \leq 1.5 \%,$ and ${\cal R}_{\chi_{\rm eff}} \leq 1.2 \%$.
Here, we only present the errors for the MPA1 EOS model but the difference from those for ARP4 is within $1\%$.

For convenience, we list the result for $\tilde{\Lambda}$ at the bottom of table \ref{tab.NS-error}, separately.
For aLIGO, the tidal deformability measurement is much worse than the measurements of the other parameters.
The errors are widely distributed as $35 \lesssim \sigma_{\tilde{\Lambda}} \lesssim 155$ depending on the mass for aLIGO,
and they can be reduced to $2 \lesssim \sigma_{\tilde{\Lambda}} \lesssim 8$ for ET, resulting in the error ratios $3.6 \% \lesssim {\cal R}_{\tilde{\Lambda}} \lesssim 6.1 \%$.
 The error difference between the two EOSs is less than $5 \ (1) \%$ for aLIGO (ET) for the binary $m_i=2\msun$,
but it becomes larger for lower masses, giving $\sim 30 \ (9) \%$ for the binary $m_i=1\msun$.
Unlike the intrinsic parameters, the measurement accuracy of the tidal parameter has a strong dependence on the SNR (the SNRs are given in the last column in the upper two subtables).
This is because, most tidal effects occur above several hundred Hz, so the measurement is almost unaffected by the entire wave cycles.
Thus, the measurement of the tidal parameter depends on the SNR rather than the binary mass that governs the entire wave cycles.

\begin{figure}[t]
           \includegraphics[width=\columnwidth]{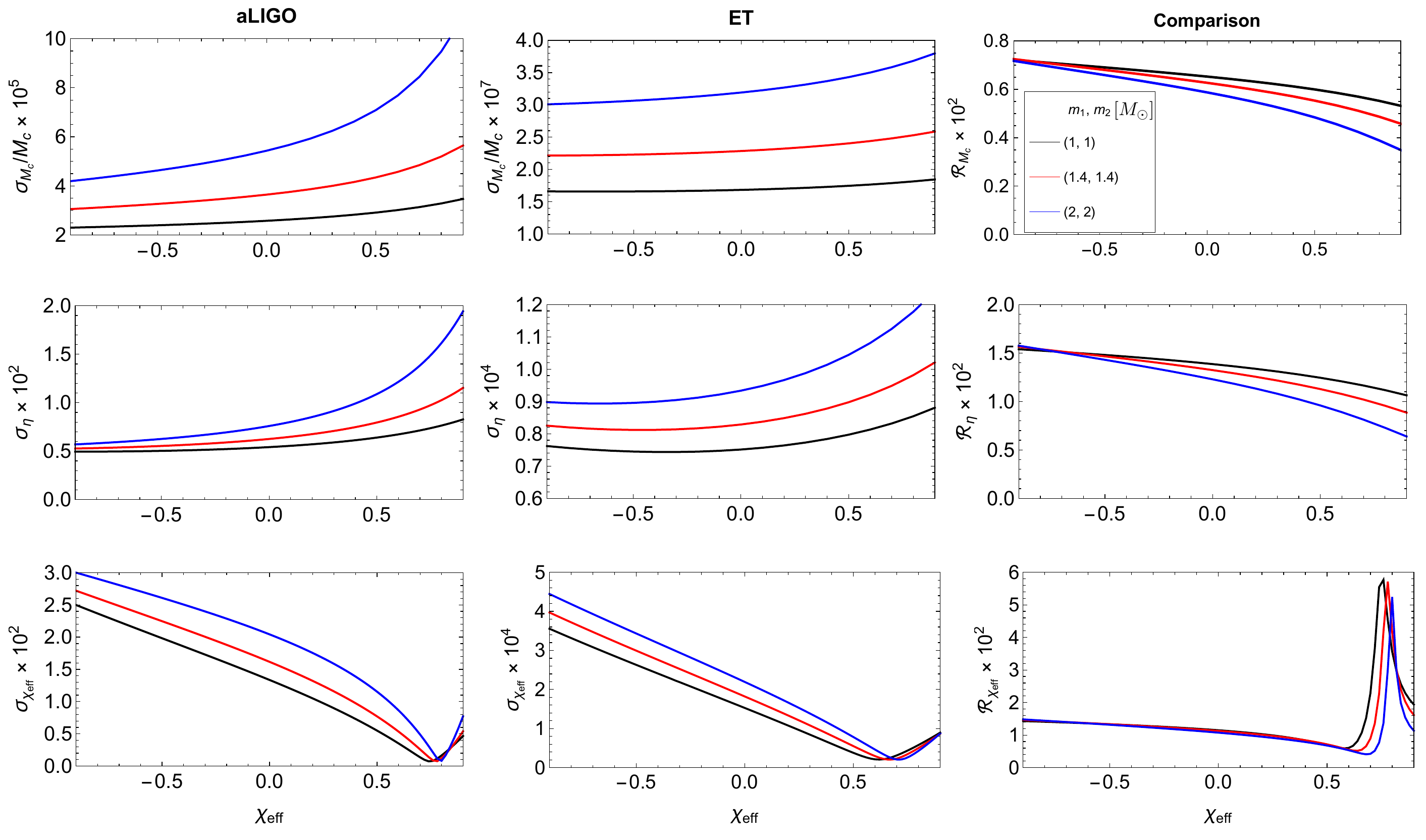}  
                  \caption{\label{fig.BNS-errors}Comparison of parameter measurement errors  of the chirp mass (top), the symmetric mass ratio (middle), and the effective spin (bottom) between aLIGO and ET for BNSs. The error ratio is defined by $ {{\cal R}_{\lambda}} = \sigma_{\lambda,\rm{ET}} /  \sigma_{\lambda,\rm{aLIGO}}  $.
  We only present the results for the MPA1 EOS model, but the results for ARP4 are indistinguishable from them.                }
\end{figure}

\begin{figure}[t]
\begin{center}
           \includegraphics[width=0.8\columnwidth]{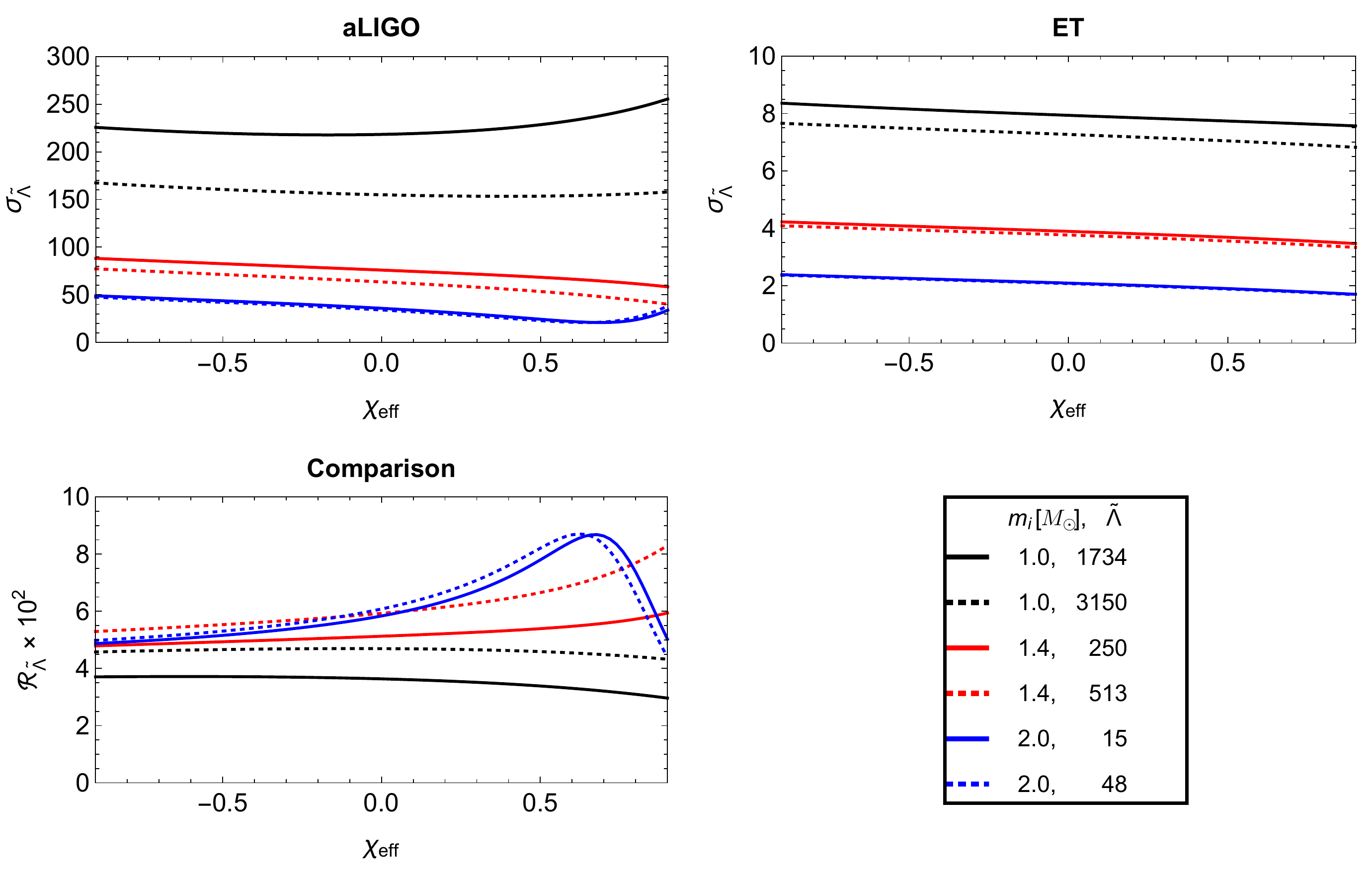}  
                  \caption{\label{fig.tidal-error}Measurement errors of the tidal deformability. The error ratio is defined by $ {{\cal R}_{\lambda}} = \sigma_{\lambda,\rm{ET}} /  \sigma_{\lambda,\rm{aLIGO}}  $. For the true values of $\tilde{\Lambda}$, we use the soft APR4 EOS model (solid line) and the stiffer EOS model MPA1 (dotted line) given in \cite{PhysRevD.79.124032}.}
                  \end{center}
\end{figure}

We display the errors as a function of $\chi_{\rm eff}$ in figure \ref{fig.BNS-errors} for the parameters $M_c, \eta,$ and $\chi_{\rm eff}$. 
Contrary to the case of BBHs, the error curves show similar trends between aLIGO and ET and relatively simple behavior.
The errors $\sigma_M/M_c$ and $\sigma_{\eta}$ monotonically  increase with increasing $\chi_{\rm eff}$, while $\sigma_{\chi_{\rm eff}}$  decreases and  increases again from $\chi_{\rm eff} \sim 0.7$.
The comparison results also show a simple behavior.
We  have the error ratios  of
${\cal R}_{M_c} \leq 0.8 \%, \ {\cal R}_{\eta} \leq 1.6 \%$,   and ${\cal R}_{\chi_{\rm eff}} \leq 1.5 \%$ exept for the high spin region.
Again, we only present the results for the MPA1 EOS model, but the results for ARP4 are indistinguishable from them.

We also show the result for $\tilde{\Lambda}$ in figure \ref{fig.tidal-error}, separately.
Overall, the variations in the errors are very small,
so the true value of $\chi_{\rm eff}$ has little effect on the measurement of $\tilde{\Lambda}$.
In our result, the maximum value of ${\cal R}_{\tilde{\Lambda}}$ is $\sim  9 \%$ at $\chi_{\rm eff} \sim 0.65$ for the binary $m_i=2\msun$.
For all BNSs, we have $\sigma_{\phi_c} \ll 2\pi$, so we do not consider the prior effect.

\section{Summary and future work} 
In this work, we demonstrated the improvement of accuracy in measuring the source parameters between aLIGO and ET
for low-mass BBHs with $M\leq 16 \msun$ and BNSs.
We calculated the Fisher matrices using $10^4$ BBH Monte-Carlo samples 
randomly distributed in our mass and spin parameter space and located at $D_{\rm eff}=400{\rm Mpc}$. 
We found that ET can detect the GW signals with $\sim 14$ times better SNRs than aLIGO
and the measurement errors of the intrinsic parameters $M_c, \eta,$ and $\chi_{\rm eff}$ for ET can be below $\sim 3 \ (7) \%$ of those for aLIGO in the range $\chi_{\rm eff}<0.3 \ (> 0.3)$.
By applying the Gaussian prior with a variance $P^2$ to the Fisher matrix,
we investigated the effect of prior on the measurement accuracy.
We found that 
the  error of the intrinsic parameters  can be reduced to $\sim 70\%$ of the original priorless error $\sigma_{\phi_c}^{\rm priorless}$
 if the prior size is $P_{\phi_c} \sim \sigma_{\phi_c}^{\rm priorless}$.
We also found that for aLIGO, the inclusion of prior information about the parameter $\phi_c$ as $P_{\phi_c}=2\pi$
can greatly reduce the error of the intrinsic parameters, but the results for ET are not affected by this prior.

We also performed the same analysis for the BNSs using equal-mass binaries with masses of $m_i=1, 1.4, 2\msun$ located at $D_{\rm eff}=40$ Mpc
and obtained the error ratios between aLIGO and ET for the intrinsic paremeters  as
${\cal R}_{M_c} \leq 0.7 \%, {\cal R}_{\eta} \leq 1.5 \%,$ and ${\cal R}_{\chi_{\rm eff}} \leq 1.2 \%$.
These results are almost independent of the EOS.
Assuming a soft and a stiffer EOS models, we calculated the errors of tidal deformability $\tilde{\Lambda}$
and found that the measurement error of $\tilde{\Lambda}$ can also be significantly reduced by ET.
We obtained the errors of $35 \lesssim \sigma_{\tilde{\Lambda}} \lesssim 155$ for aLIGO and $2 \lesssim \sigma_{\tilde{\Lambda}} \lesssim 8$ for ET
and the error ratio of $3.6 \% \lesssim {\cal R}_{\tilde{\Lambda}} \lesssim 6.1 \%$.
These precise measurements of the NS tidal parameter will play an important role in constraining various theoretical models for the NS EOS.

Although only low-mass BBHs were considered here, the masses of the observed BBHs are broadly distributed up to $\sim 160 \msun$ \cite{GWTC-1,GWTC-2}.
Since TaylorF2 is only valid for low-mass systems, one has to use the inspiral-merger-ringdown waveform models, such as the phenomenological models  \cite{Kha16,Aji11,San10},
to obtain correct measurement errors for high-mass BBHs.
However, the phenomenological models are much more complicated than TaylorF2; thus, applying those models to the Fisher matrix method would be very difficult.
Nevertheless, we will extend this work to the high-mass BBH system by using one of the recent phenomenological models
and cross-check the results in this work.
We will also use the python package {\bf GWBench} which was recently developed by \cite{Borhanian_2021} designed specifically for GW Fisher Matrix analyses.
GWBench supports the IMRPhenom models via numerical differentiation methods, we will check the consistency between GWBench and the analytical method.
Some results for high-mass BBHs are given in \ref{ap.high-mass-bbh} obtained by using GWBench.

\ack{This work was supported by the National Research
Foundation of Korea (NRF) grants funded by the Korea
government (No. 2018R1D1A1B07048599, No. 2019R1I1A2A01041244, and No. 2020R1C1C1003250}


\appendix

\section{Effect of the extrinsic parameters on the measurement errors of the intrinsic parameters}  \label{ap.extrinsic}
The effective distance is given by \cite{All12}
\be   \label{eq.deff}   \label{eq.Deff}
D_{\rm eff}=D / \sqrt{F^2_+ \bigg(\frac {1+\cos^2 \iota}{2}   \bigg)^2 + F^2_{\times}\cos^2 \iota} ,
\ee
where $D$ is the true distance of the source, $F_+$ and $F_{\times}$ are the antenna response functions for the incident signal; 
these functions can be given by the sky position (RA, Dec) of the source with respect to the reference frame of the detector and the polirization angle $\psi$.
If the detector is placed at the x--y plane and the detector arms are aligned with the axes, 
the sky position is given by $(\theta, \phi)$ in the spherical coordinate system,
and the antenna response functions are given by
\bea
F_+=\frac{1}{2} \cos 2 \phi  \cos 2 \psi  \left(\cos ^2\theta +1\right)-\sin 2 \phi  \sin 2 \psi  \cos \theta ,\\
F_{\times}=   \frac{1}{2} \cos 2 \phi  \sin 2 \psi  \left(\cos ^2\theta +1\right)+\sin 2 \phi \cos 2 \psi \cos (\theta ).
  \eea
For a fixed $D$, the  maximum SNR (or minimum $D_{\rm eff}$) can be obtained when
$\theta=\phi=\iota=0$.
In this case, $D_{\rm eff}$ is the same as $D$ and independent of $\psi$.

On the other hand, different true values for $\theta, \phi$, and $\iota$ give different SNRs, 
so all measurement errors of the intrinsic parameters are adjusted simultaneously by the new SNR. 
In order to see the effect of the extrinsic parameters on the measurement of the intrinsic parameters,
we investigate the variation of SNR by changing the extrinsic parameters.
It is straightforward that the SNR is inversely proportional to $D$ for a given sky position and inclination.
At the optimal sky position ($\theta=\phi=0$) with $\psi=0$, the effective distance can be only a function of inclination:
\be
D_{\rm eff}=\frac{2 D}{\cos ^2 {\iota }+1},
\ee
which gives, max($D_{\rm eff})=2D$ at $\iota=\pi/2$, and min($D_{\rm eff})=D$ at $\iota=0$  or $\pi$.
The strength of the incident signal is independent of the polarization angle $\phi$ because that parameter is only determined by the polarization basis for a fixed source. 
On the other hand, assuming $\iota=0$ in equation (\ref{eq.Deff}), we have
$D_{\rm eff}=D/\sqrt{F^2_++F^2_{\times}}$, so $\rho^2 \propto D_{\rm eff}^{-2} \propto F^2_++F^2_{\times}$, and the function
\be
F^2_++F^2_{\times}=\frac{1}{4} \cos ^22 \phi  \left(\cos ^2\theta +1\right)^2+\sin ^22 \phi  \cos ^2\theta 
\ee
corresponds to the antenna power pattern of a single detector, which is independent of $\psi$.
This is often referred to as the ``peanut diagram".
If the signal comes from the sky position $\theta=\pi/2$ and  $\phi=(1+2n)\pi/4$, 
 $D_{\rm eff} \rightarrow \infty$. Therefore, there can be four hidden positions for a single detector observation.

The measurement error of $D_{\rm eff}$ can also be calculated by including that parameter in the Fisher matrix,
although $D_{\rm eff}$ is mostly uncorrelated with the intrinsic parameters. 
A comparison between a 5-dimensional Fisher matrix neglecting $D_{\rm eff}$ and a 6-dimensional Fisher marix including $D_{\rm eff}$
is shown in table \ref{tab.5dvs6d}.
On the other hand, one single extrinsic parameter incorporated in the function of $D_{\rm eff}$ can also be used in the Fisher matrix,
 where the true values of the other four parameters should be set to be fixed.
However, if we choose $\iota$ as the variable, not all angles are amenable to the Fisher Matrix approximation.
For example, if $\iota=n \pi/2$, then $\partial{h}/\partial{\iota}=0$, and the Fisher matrix becomes degenerate.

\begin{table}[t]
\begin{center}
\caption{\label{tab.5dvs6d}{Comparison of the errors ($\sigma$) between 3, 5, and 6-dimensional Fisher matrices.  The true values are assmumed to be  ($m_1, m_2, \chi_{\rm eff}, t_c, \phi_c, D_{\rm eff} )=(10\msun, 5\msun, -0.5, 0, 0, 400 {\rm Mpc}$). In the 6-dimensional Fisher matrix result, the correlations between $D_{\rm eff}$ and the intrinsic parameters  are given as ($\cal{C}_{\it M_c D_{\rm eff}},  \cal{C}_{\it \eta D_{\rm eff}},  \cal{C}_{\it \chi_{\rm eff} D_{\rm eff}}$)=(0.0295, -0.0288, 0.0289). We use the aLIGO PSD and $f_{\rm low}=10$Hz.}}
\begin{indented}
\item[]\begin{tabular}{  c cccccc}
\br
Dimension&$\sigma_{M_c }/M_c {\times} {10^3}$&$\sigma_{\eta}$&$\sigma_{\chi_{\rm eff}}$ & $t_c [{\rm s}]$& $\phi_c$& $\sigma_{D_{\rm eff}}$  \\
\mr
3-D &0.1093&0.005416&0.01896&-&-&-     \\
5-D  &1.0712&0.036853 & 0.18528 &0.0015430 &8.648& - \\
6-D   & 1.0716 &0.036868 & 0.18539 & 0.0015436&8.652&12.1 \\
\br
\end{tabular} 
\end{indented}
\end{center}
\end{table}

\section{Comparison  between the Fisher matrix and the Bayesian parameter estimation posteriors} \label{ap.bilby}
\begin{figure}[t]
\begin{center}
           \includegraphics[width=0.9\columnwidth]{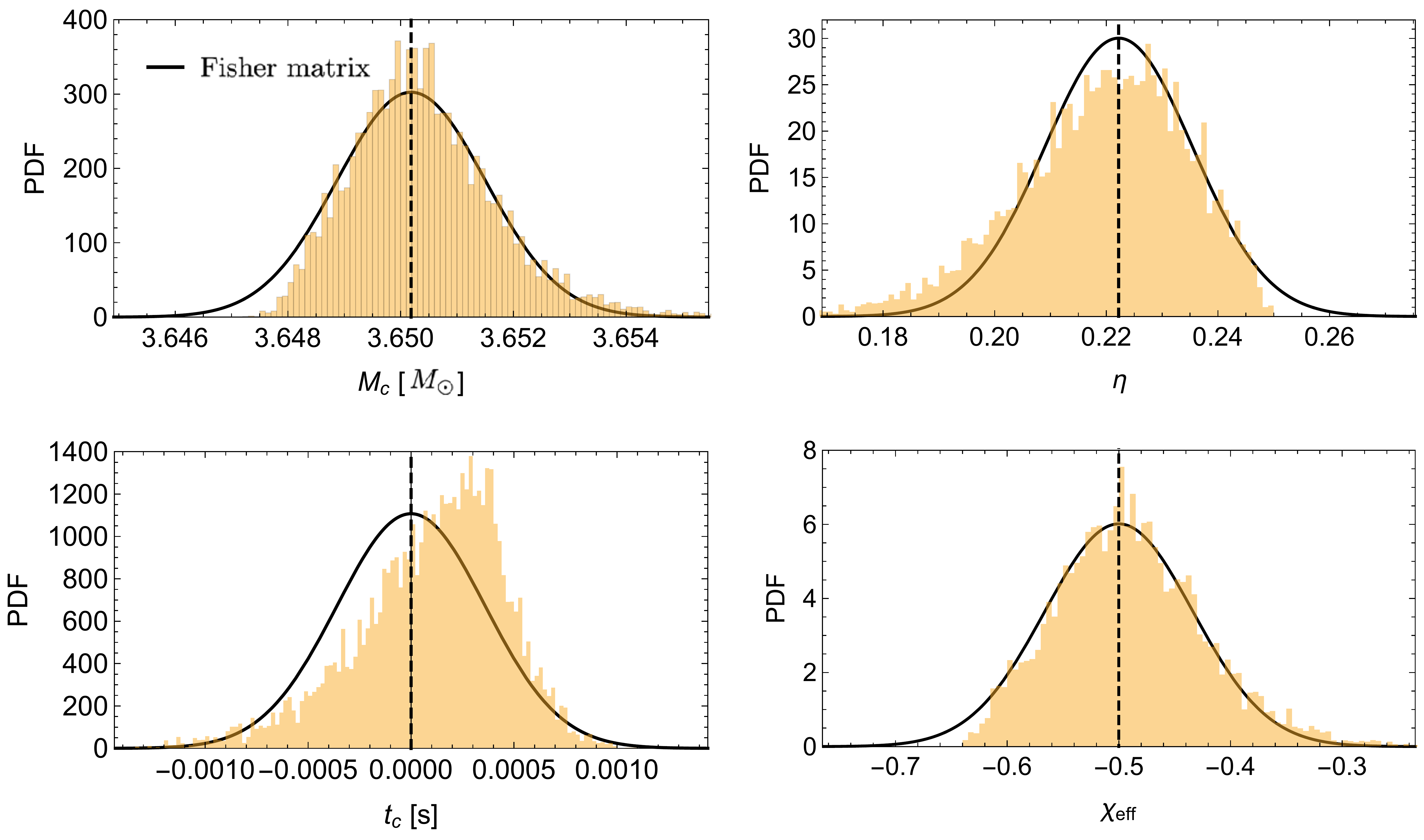}  
                  \caption{\label{fig.FM-bilby-comparison} Comparison between the Fisher matrix and the Bayesian parameter estimation posteriors for a BBH. We assume the true values to be $(m_1, m_2, \chi_1, \chi_2, t_c, \phi_c)=(6\msun, 3\msun, -0.5, -0.5,0, 0)$ and use the  aLIGO PSD with $f_{\rm low}=20$. The SNR is $100$.}
                  \end{center}
\end{figure}

We perform Monte-Carlo parameter estimation simulations using the Bayesian parameter estimation package {\bf Bilby} \cite{Ashton_2019},
which is a user-friendly Bayesian inference library.
We use one of the packaged samplers, {\bf Dynesty}\footnote{https://github.com/joshspeagle/dynesty}.
As a fiducail binary, we adopt a BBH with the true parameter values of $(m_1, m_2, \chi_1, \chi_2,t_c, \phi_c)=(6\msun, 3\msun, -0.5, -0.5, 0, 0)$,
and we use the aLIGO PSD assuming $f_{\rm low}=20$Hz.
Our Fisher matrix is 5-dimensional since we use $\chi_{\rm eff}$ instead of $\chi_1, \chi_2$, while the Monte-Carlo simulations run in 6-dimensional space
with fixed extrinsic parameters.
Since we choose sufficiently high SNR ($\rho=100$), we do not need to consider the effect of priors on our Fisher matrix result.
In the Bayesian parameter estimation, the priors are given as flat distributions within physical boundaries, e.g. $[0, 2\pi]$ for $\phi_c$.
In figure \ref{fig.FM-bilby-comparison}, we show a comparison between the Fisher matrix and the Bayesian parameter estimation posteriors.
One can see that our Fisher matrix result agrees remarkably well with the Bayesian posteriors.
In Bilby, one can choose the parameters when displaying the Bayesian posterior PDFs, so we use $\chi_{\rm eff}$ for direct comparison with the Fisher matrix result.
On the other hand, the Bayesian posterior of $\phi_c$ cannot be given because we have selected the phase marginalization option for efficiency.

\section{Comparison between the optimal SNR and the matched-filter SNR}  \label{ap.snr-comparison}
\begin{figure}[t]
\begin{center}
           \includegraphics[width=0.8
          \columnwidth]{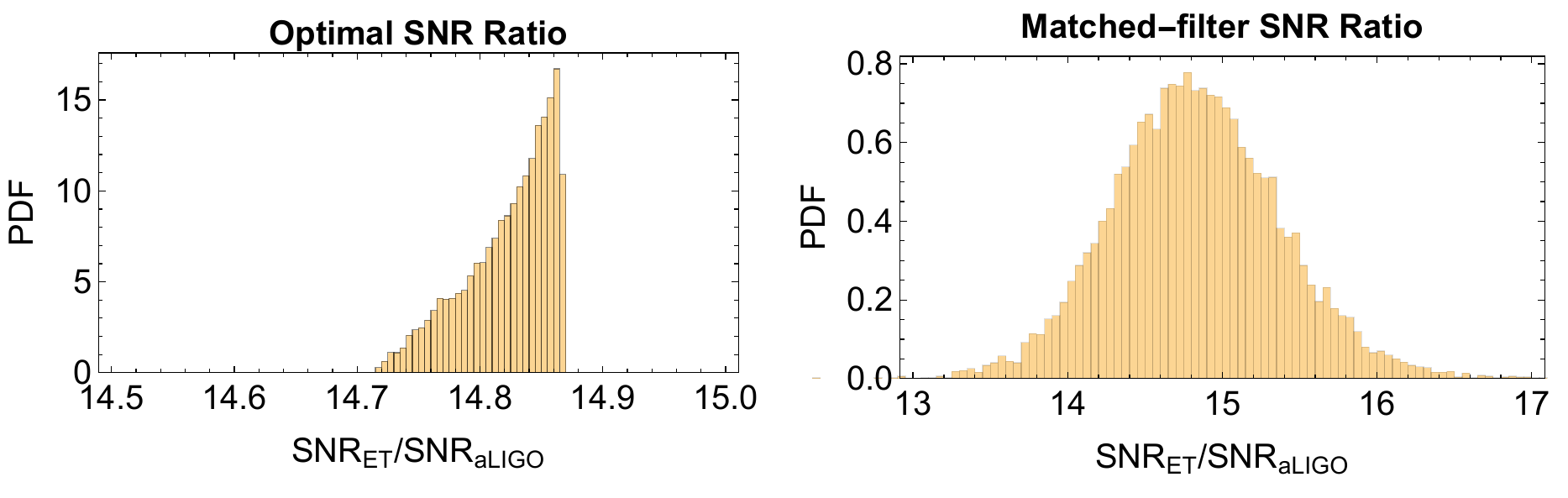}  
                  \caption{\label{fig.opt-match-snr} The optimal SNR ratio (left) and the matched-filter SNR ratio (right) between aLIGO and ET.
                  We assume $f_{\rm low}=20 \ (10)$ Hz for aLIGO (ET).}
                  \end{center}
\end{figure}

We perform the injection studies to have more realistic SNRs using Bilby.
We inject $10^4$ BBH signals into Gaussian noise using the aLIGO and ET PSDs, respectively, and compute the matched-filter SNRs.
The true parameter values are randomly distributed in our 3-dimensional mass and spin parameter space with a fixed $D_{\rm eff}$.
The matched-filter SNR is given by $<s+n|s>/\rho$, where $n$ is the Gaussian noise and $\rho$ is the optimal SNR defined in equation (\ref{eq.snr}). 
Thus, if we remove the Gaussian noise in this computation,  the matched-filter SNR can be the same as the optimal SNR.
We compute both the optimal SNR and the matched-filter SNR for aLIGO and ET and
calculate the SNR ratio between the two detectors.
Figure \ref{fig.opt-match-snr} shows a comparison between the optimal SNR ratio (${\cal R}_{\rm SNR}^{\rm optimal}$) and the matched-filter SNR ratio (${\cal R}_{\rm SNR}^{\rm search}$).
One can see that the distribution of ${\cal R}_{\rm SNR}^{\rm optimal}$ is very narrow $\sim [14.7, 14.9]$ and shows a peak near the highest region, 
while ${\cal R}_{\rm SNR}^{\rm search}$ shows a much wider and Gaussian-like distribution, centered at the same position as the peak position of ${\cal R}_{\rm SNR}^{\rm optimal}$.
Here, we assume $f_{\rm low}=20 \ (10)$ Hz for aLIGO (ET) due to difficulties in setting the initial frequency below 20 (10) Hz in Bilby, while
we used $f_{\rm low}=10 \ (1)$ for aLIGO (ET)  in our analyses in section \ref{sec.result}.
However, we found that the difference between the SNRs for $f_{\rm low}=10 \ (1)$ Hz and $f_{\rm low}=20\ (10)$ Hz is within $\sim 2 \ (0.5)\%$ for aLIGO (ET).
Because of these small differences, ${\cal R}_{\rm SNR}^{\rm optimal}$ in this figure is slightly larger than ${\cal R}_{\rm SNR}$ in table \ref{tab.intrinsic-errors}.
 
\section{Multi-detector SNR analysis for the second and third-generation detectors} \label{ap.multi-detector}
The advantage of multi-detector networks in the measurement accuracy of intrinsic parameters can be achieved mainly by an increase in the SNR.
Here, we investigate the single and the network SNRs for the second and third-generation detectors.
For the second-generation detectors we consider aLIGO Hanford (H) and Livingstone (L) \cite{Aasi_2015}, advanced Virgo (V) \cite{AVirgo}, and KAGRA (K) \cite{PhysRevD.88.043007}.
For  third-generation detectors, we consider the Einstein Telescope (E) \cite{Punturo_2010} and the Cosmic Explorer (C) \cite{2017CQGra..34d4001A}.
For ET, we use the triangular detector configuration consisting of the three sub-detectors E1, E2, and E3.
Bilby includes geographic coordinates, orientations, and PSDs for all of the detectors above.
We inject $10^4$ BBH sources distributed over the sky position
and calculate the matched-filter SNRs for the detectors.
In this calculations, we randomly generate the angle RA in the range $[0,2\pi]$ and $\cos ({\rm Dec})$ in the range $[-1,1]$.
We do not take into account the binary orientation by choosing $\iota=0, \psi=0,$ and $D=400$ Mpc for all samples. 
As a fiducial binary, we choose a nonspinning BBH with the masses $(6\msun, 3\msun)$.
We assume $f_{\rm low}=20$ H and  $10$ Hz for the second and the third-generation detectors, respectively.

\begin{figure}[t]
           \includegraphics[width=\columnwidth]{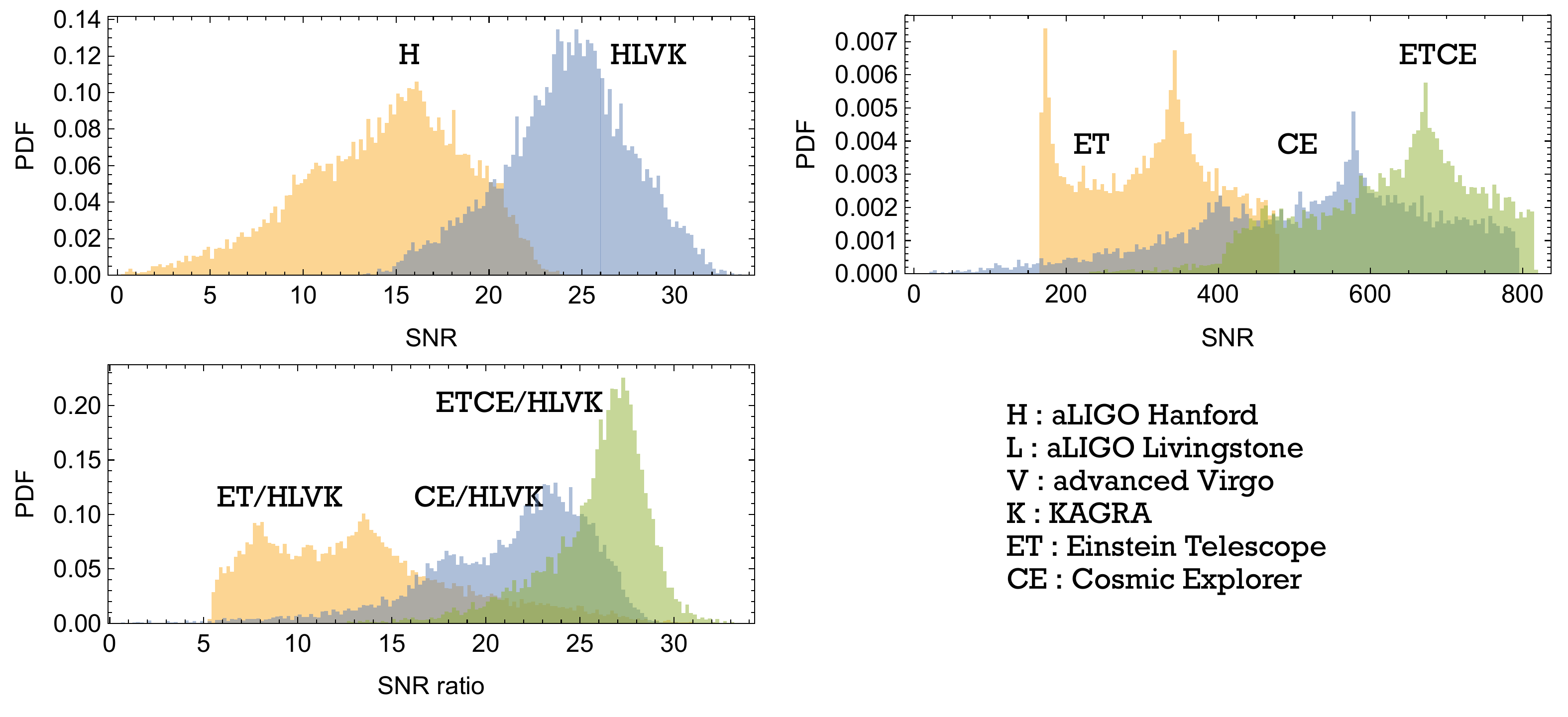}  
                  \caption{\label{fig.network-snr} PDFs for the single and network SNRs for the second-generation (upper left panel) and the third-generation (upper right panel) detectors and their SNR ratios (lower panel). As a ficucial binary, we assume a nonspinning BBH with the masses $(6\msun, 3\msun)$. We assume $f_{\rm low}=20$ H and  $10$ Hz for the second-generation detectors and the third-generation detectors, respectively.}
\end{figure}

The result is given in figure \ref{fig.network-snr}.
The upper-left panel shows the single SNRs for H and the network SNRs for the second-generation detectors,
and the upper-right panel shows the single and the network SNRs for the third-generation detectors.
Note that, since ET consists of the three sub-detectors, a single SNR for ET corresponds to the network SNR for the three sub-detectors.
Finally, the bottom panel shows the ratios between the second and the third-generation network SNRs assuming various combinations.

\section{Subtleties in the inversion of the Fisher matrix} \label{ap.inversion}
In this work, we evaluate the Fisher matrix and its inversion using a Mathematica code.
Following the methodology of \cite{PhysRevD.71.084025}, we check the reliability of the result simply by multiplying the inversion by the original matrix.
We define the numerical identity matrix as $I^{\rm num}$ and measure the deviation from the true indentity matrix:
\be
\epsilon_{\rm inv}\equiv {\rm max}_{ij} | I^{\rm num}_{ij}-\delta_{ij}|,
\ee
where $\delta_{ij}$ is the Kronecker symbol.
We find that  all of the BBH samples used in our 5-dimensional Fisher matrices
could achieve $\epsilon_{\rm inv} < 10^{-6}$,
and $\epsilon_{\rm inv} < 10^{-5}$ for the 6-dimensional Fisher matrices for BNSs (cf. Appendix B of \cite{PhysRevD.71.084025}).

\section{High-mass BBHs} \label{ap.high-mass-bbh}
\begin{figure}[t]
           \includegraphics[width=\columnwidth]{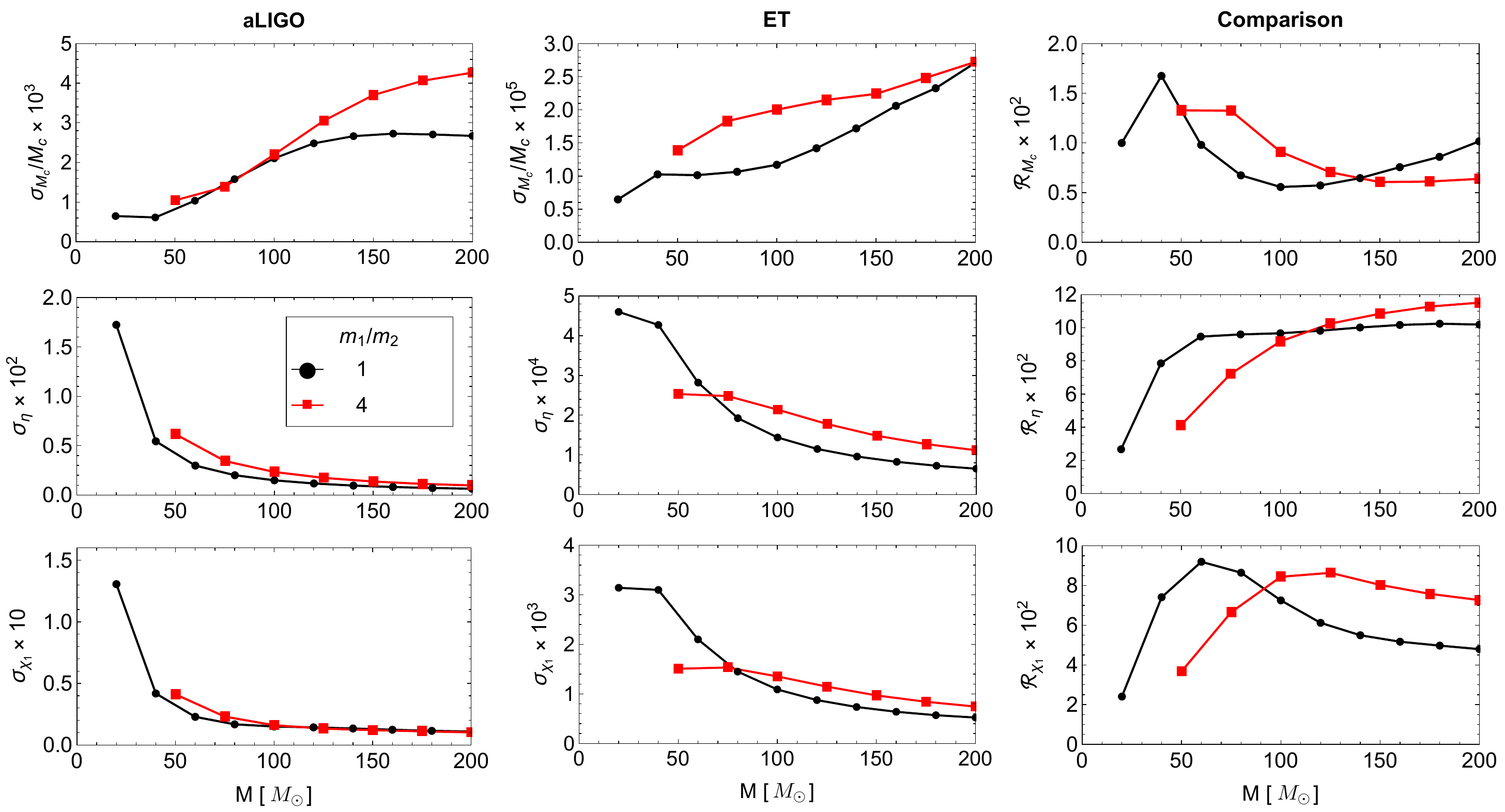}  
                  \caption{\label{fig.high-mass-BBH} One-dimensional error curves for high-mass BBHs given as a function of the total mass  with fixed mass ratios of $m_1/m_2=1$ and 4. We assume $\chi_1=\chi_2=-0.1, D_{\rm eff}=400$ Mpc, and $f_{\rm low}=10$  and  $1$ Hz for aLIGO and ET, respectively.}
\end{figure}

\begin{figure}[t]
           \includegraphics[width=\columnwidth]{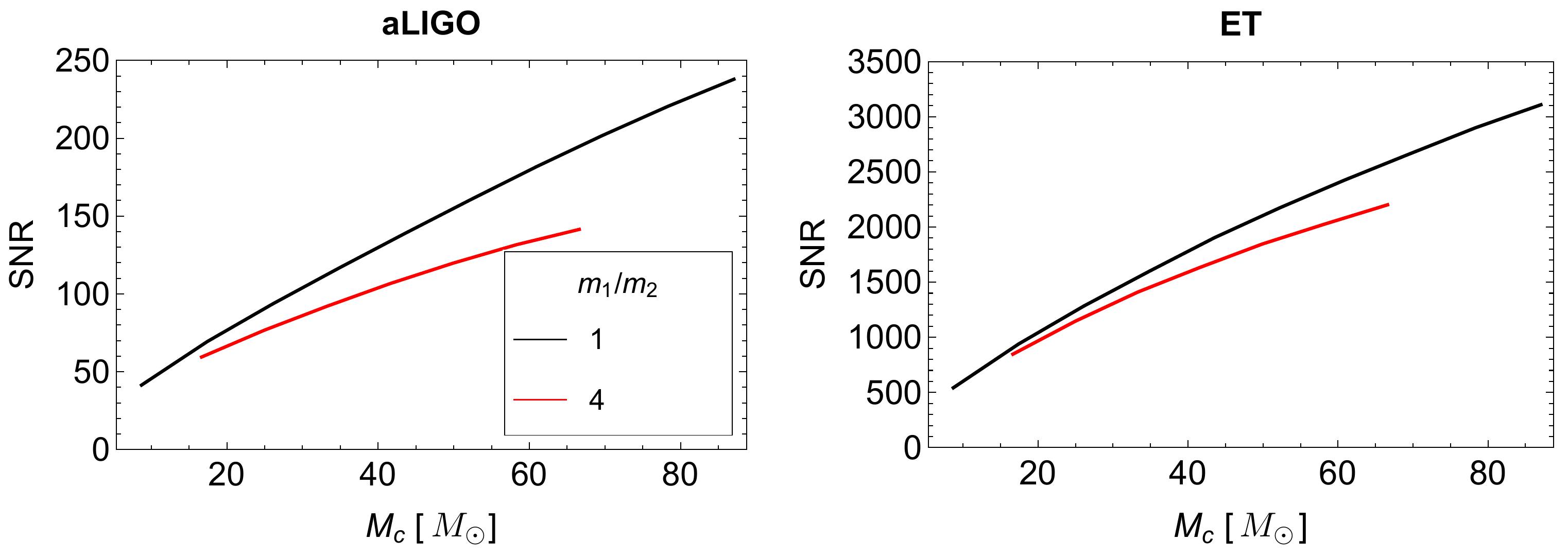}  
                  \caption{\label{fig.high-mass-BBH-SNR} The optimal SNRs for high-mass BBHs. The SNR ratio is $13 \lesssim {\cal R}_{\rm SNR}^{\rm optimal}  \lesssim 16$.}
\end{figure}

Here, we show some results for high-mass BBHs obtained by using GWBench.
We employ IMRPhenomHM \cite{PhysRevLett.120.161102} which is a higher-multipole model based on IMRPhenomD,
this model is more accurate than dominant-multipole-only models, especially for highly-asymmetric binaries.
Figure \ref{fig.high-mass-BBH} shows the errors as a function of the total mass
assuming symmetric and asymmetric-mass binaries with $m_1/m_2=1$ and 4.
Since we fix the spin as $\chi_1=\chi_2=-0.1$, this result does not show any spin dependence.
We will perform a more comprehensive study for high-mass BBHs in a broad parameter space in future work.
Since only the component spins can be used in GWbench, we only use the spin parameter $\chi_1$ so the Fisher matrix can be given by $5 \times 5$ matrix with the parameters
$(M_c, \eta, \chi_1, t_c, \phi_c)$.  The choice of the spin parameter between $\chi_1$ or $\chi_{\rm eff}$ has a little impact on the measurement errors of the parameters considered.
Note that, using both component spins in the Fisher matrix can significantly increase the errors due to increasing parameter dimension.
Figure \ref{fig.high-mass-BBH-SNR} shows the SNRs as a function of the chirp mass. Unlike the result for small-mass BBHs in figure \ref{fig.snr}, the two curves are clearly distinct.
That means, the dependence of the SNR on the mass ratio becomes stronger for higher-mass systems
due to the non-negligible contribution of the merger-ringdown.  The SNR ratio between aLIGO and ET is given as $13 \lesssim {\cal R}_{\rm SNR}^{\rm optimal}  \lesssim 16$.

%
%
%
\section*{References}

\bibliographystyle{aip}
\bibliography{biblio}

\end{document}